\documentclass[openacc]{rstransa}%%%%where rstrans is the template name

\usepackage{natbib}
\usepackage{graphicx}
\usepackage{amsmath}

\titlehead{Research}

\begin{document}

%%%% Article title to be placed here
\title{The influence of thermal pressure gradients and ionization (im)balance on the ambipolar diffusion and charge-neutral drifts}

\author{%%%% Author details
 M.M. G\'omez M{\'\i}guez$^{1,2}$, D. Mart{\'\i}nez G\'omez$^{1,2}$, E. Khomenko$^{1,2}$, N. Vitas$^{1,2}$}

%%%%%%%%% Insert author address here
\address{$^{1}$ Instituto de Astrof\'{\i}sica de Canarias, 38205 La Laguna, Tenerife, Spain \\
$^{2}$ Departamento de Astrof\'{\i}sica, Universidad de La Laguna, 38205, La Laguna, Tenerife, Spain}

%%%% Subject entries to be placed here %%%%
\subject{Astrophysics, Solar System, Space Exploration, Sun}

%%%% Keyword entries to be placed here %%%%

%%%% Insert corresponding author and its email address}
\corres{M.M. Gómez-Míguez\\
\email{martin.manuel.gomez.miguez@iac.es}}

%%%% Abstract text to be placed here %%%%%%%%%%%%
\begin{abstract}
Solar partially ionized plasma is frequently modeled using single-fluid (1F) or two-fluid (2F) approaches. In the 1F case, charge-neutral interactions are often described through ambipolar diffusion, while the 2F model fully considers charge-neutral drifts. Here, we expand the definition of the ambipolar diffusion coefficient to include inelastic collisions (ion/rec) in two cases: a VAL3C 1D model and a 2F simulations of the Rayleigh-Taylor instability (RTI) in a solar prominence thread based on \cite{PopLukKho2021aa, PopLukKho2021ab}. On one side, we evaluate the relative importance of the inelastic contribution, compared to elastic and charge-exchange collisions.  On the other side, we compare the contributions of ion/rec, thermal pressure, viscosity, and magnetic forces to the charge-neutral drift velocity of the turbulent flow of the RTI. Our analysis reveals that the contribution of inelastic collisions to the ambipolar diffusion coefficient is negligible across the chromosphere, allowing the classical definition of this coefficient to be safely used in 1F modeling. However, in the transition region, the contribution of inelastic collisions can become as significant as that of elastic collisions. Furthermore, we ascertain that the thermal pressure force predominantly influences the charge-neutral drifts in the RTI model, surpassing the impact of the magnetic force.

\end{abstract}%%%%%%%%%%%%%%%%%%%%%%%%%%%

\maketitle
%%%%%%%%%% Insert the texts which can accomdate on firstpage in the tag "fmtext" %%%%%

%\begin{fmtext}
\section{Introduction\label{sec:1}}

Solar atmospheric plasma is mainly composed of hydrogen ($\sim 73.5 \%$) and helium ($\sim 25 \%$), with small traces of  heavier elements such as oxygen, carbon, iron, and neon \cite{VerAvrLoe1981}. Roughly estimating the ionization fraction from Saha's equation in solar conditions, it follows that hydrogen is almost fully ionized for temperatures above $10^4 \ \rm{K}$ \cite{Sah1921aa}. The solar corona, with temperatures of the order of $10^{6} \ \rm{K}$, is expected to be fully ionized, while the solar chromosphere and photosphere, with temperatures in the range between $10^{3}$ and $10^{4} \ \rm{K}$, are expected to be partially ionized.

Although a plasma can be composed of many species with very distinct physical properties such as ions, electrons, or neutrals of different chemical elements, for phenomena with characteristic temporal scales longer than collision time scales between all its components, the plasma behaves as a single fluid \cite[see e.g.,][]{Bra1965aa,KhoColDia2014aa,BalAleCol2018aa}. The most extended model for the description of the fully collisionally coupled solar plasma is the ideal magnetohydrodynamics (ideal MHD). It succeeds in explaining large-scale phenomena such as magneto-convection, waves, sunspot structure, and dynamics, among others, \cite{Pri1982aa,GoePoe2004aa}. Despite its achievements, additional physics needs to be taken into account to understand phenomena at smaller temporal and spatial scales. 

The most popular extensions of the ideal MHD for partially ionized plasmas are the so-called single-fluid (1F), two-fluid (2F), or multi-fluid (MF) approaches. The 1F approach considers all the plasma components moving together with a single center of mass velocity, but allowing for small drifts among the components. In this case, the generalized Ohm's law relates the electric and magnetic field through the center of mass velocity of the whole plasma, which deviates from the velocities of charges, leading to non-ideal effects \cite{MarDePHan2012aa, KhoCol2012aa, BalAleCol2018aa, PopKep2021aa}. In the 2F approach, frequently, but not exclusively, the charged particles (ions and electrons) are grouped in a single fluid, and all neutral components form another fluid. Now, the generalized Ohm's law naturally uses the center of mass velocity of charges as a reference. From this one, the expression for the 1F case can be derived by changing the system of reference for velocity \cite{Bra1965aa, SchNag2009aa, Mei2011aa, KhoColDia2014aa, BalAleCol2018aa}. A direct evaluation of the conditions of the solar upper photosphere and chromosphere, suggests that the leading term in the 1F Ohm's law is the ambipolar diffusion, followed by the modified Hall effect and Biermann battery effect \cite{MarDePHan2012aa, KhoCol2012aa, KhoColDia2014aa, KhoVitCol2018aa, GonKhoVit2020aa}. The 1F drift velocity, obtained by combining the 2F momentum equations for charges and neutrals exhibits contributions from the magnetic force, thermal pressure gradient force, and a smaller contribution from the ion-electron drift, proportional to the electron mass \cite{Bra1965aa, ForOliBal2007aa,KhoColDia2014aa}. The latter is frequently neglected together with the force derived from ionization/recombination.

Many authors include non-ideal effects derived from the presence of neutrals into their analysis to account for heating or multi-scale physics in the multi-fluid approximation \cite[see e.g.,][to name a few]{Mei2011aa, LeaLukLin2012aa, KhoColDia2014aa, BalAleCol2018aa, PopLukKho2021aa, SnoDruHil2023aa, WarMarHan2023aa, MarDePHan2023aa, PopKep2023aa, PopLukKho2023aa}. Several of those works additionally consider other non-ideal effects such as neutral viscosity, heat flux, or thin radiative losses. In an ongoing debate, \cite{VraKrs2013aa} and \cite{AlhBalFed2022aa} questioned the MHD approach for describing magnetic field-related wave modes (fast magneto-acoustic modes or Alfv\'en modes) in the photosphere of the quiet Sun due to extremely low ionization fraction. In \cite{AlhBalFed2022aa} the authors concluded that a three-fluid model (for electrons, ions, and neutrals) has to be applied in the photosphere, and that the medium would only support acoustic modes in each fluid. Contrary to that, \cite{SolCarBal2013ab} showed that in a strongly collisionally-coupled photosphere the propagation of Alfv\'en waves is supported by both, neutral and charged fluids.

In this work, we focus on the interaction between neutrals and charges (ions + electrons), by both elastic and inelastic collisions. In the first group, neutrals and charges are modeled to collide as hard spheres. Resonant charge exchange (CX) processes are included as elastic collisions as well. In the second group, radiative and collisional ionization/recombination processes are accounted for. The expression for the ambipolar diffusion coefficient frequently omits the contribution from inelastic collisions. Hence, one of the motivations of the present work is to provide more general expressions for both the ambipolar coefficient and the 1F drift velocity, including the inelastic contributions and to evaluate their importance in the typical conditions of the solar chromosphere and prominence plasma. Inelastic collisions, alongside the elastic ones can exchange momentum between neutrals and charges. Therefore, they should be able to modify the relative velocity of the species. In many solar applications, CX collisions are also omitted, considering only the hard sphere collisions. However, omitting CX interaction in hydrogen plasmas is not well justified since it is the dominant collisional momentum transfer mechanism \cite{KrsSch1999aa,Ger2004aa}. For example, \cite{TerSolOli2015aa} in their study of neutral support against gravity in solar prominences, found that CX collisions were three times more efficient than the hard sphere collisions in keeping the neutrals from falling.

Previous studies have incorporated inelastic collisions to examine various dynamical processes within partially ionized solar plasmas, encompassing phenomena like waves, shocks, instabilities, and reconnection. The authors of \cite{ManAlvLan2017aa} were the first to include ionization/recombination processes into the modeling of propagation of magneto-acoustic waves through the gravitationally stratified solar atmosphere. Their main conclusion was that ionization/recombination imbalance affects the initial thermodynamic structure of the atmosphere, resulting in a slight modification of the wave periods with height. Similarly, in line with \cite{ManAlvLan2017aa}, \cite{ZhaPoeLan2021aa} observed that the initial equilibrium undergoes modification when chemical equilibrium is considered. This alteration leads to a more pronounced decoupling between charges and neutrals. Conversely, when the initial equilibrium is fixed, \cite{Bal2019aa} demonstrated that ionization imbalance tends to increase the collisional coupling between charges and neutrals for slow magneto-acoustic and Alfvén waves. Moreover, \cite{MurHilSno2022aa} discovered that due to effectively enhanced coupling, ionization-recombination processes stabilize current sheets and dampen their dynamics, subsequently reducing reconnection rates during the coalescence instability. Accounting for inelastic processes was shown to non-trivially change the substructure of shocks, and post-shock regions, therefore affecting the charge-neutral decoupling over the shock fronts \cite{SnoHil2021aa}. Ionization imbalance was also included in the simulations of Rayleigh-Taylor instability (RTI) in \cite{PopLukKho2021aa, PopLukKho2021ab}, revealing the creation of an over-ionized layer surrounding falling drops. However, aside from this effect, no significant influence on charge-neutral decoupling was observed.

Apart from the influence of the inelastic processes into the decoupling, we are also interested in evaluating the relevance of the thermal pressure function, which accounts for the differences in the pressure gradients of the charged and neutral fluids. Although this term has already been formally included in previous derivations of 1F models for partially ionized plasmas \cite{Bra1965aa,ForOliBal2007aa,PanWar2008aa,KhoColDia2014aa}, it has been usually neglected in applications to solar plasmas. Nonetheless, some works such as \cite{ForOliBal2008aa,BalCarSol2017aa} have retained it in studies of oscillations in prominences. As mentioned by \cite{Bra1965aa} and \cite{Bal2014aa}, the thermal pressure function can be safely neglected in both the limits of weakly and strongly ionized plasmas and in cold plasmas, with its effect being larger for the intermediate range of ionization fraction. More specifically, the direct calculation of \cite{KhoColDia2014aa}, using a simulated realistic model of solar magneto-convection, and a semi-empirical model of a sunspot, showed that the contribution of the thermal pressure term to the generalized Ohm's law is negligible in comparison to other processes such as ambipolar diffusion and Hall's effect. Unlike that, \cite{PopLukKho2021ab} studied the force balance in the 2F simulations of the Rayleigh-Taylor instability (RTI) at a smooth transition between solar prominence thread and corona, and noticed the large importance of the pressure gradients in balancing the charge-neutral friction force. Here, we are going to investigate the same scenario as \cite{PopLukKho2021aa,PopLukKho2021ab} but we are going to focus on the impact of the thermal pressure function on the drift velocities from the perspective of a single-fluid model.

This paper is organized as follows. In Section \ref{sec:2}, we introduce the motion equations and all the elastic and inelastic collisions considered. In Section \ref{sec:3}, the expression for the charge-neutral drift velocity, including both elastic and inelastic collisions, is obtained. The importance of each contribution to the drift velocity is evaluated under two different conditions: in Section \ref{sec:4}, we explore a model of the quiet Sun in statistical equilibrium (SE), and in Section \ref{sec:5} we analyse two 2F simulations of the RTI in a prominence thread which include ionization/recombination imbalance (this will be generically referred to non-equilibrium, or NEQ, through the paper). Finally, the conclusions are presented in Section \ref{sec:6}. 

%%%%%%%%%%%%%%%%%%%%%%%%%%%%%%%%%%%%%%%%%%%%%%%%%%%%%%%%%%%%%
\section{2F hydrogen plasma model}
\label{sec:2}
%%%%%%%%%%%%%%%%%%%%%%%%%%%%%%%%%%%%%%%%%%%%%%%%%%%%%%%%%%%%%

We start our discussion by presenting the reference model for our plasma description. For simplicity, we restrict ourselves to a pure hydrogen plasma, although it can be straightforward to extend it to a more general case. The discussion follows from the set of equations of mass balance \cite{KhoColDia2014aa, PopLukKho2019aa}
\begin{subequations}
    \begin{gather}
        \frac{\partial \rho_{\rm{n}}}{\partial t} + \nabla \cdot (\rho_{\rm{n}} \vec{u}_{\rm{n}}) = S_{\rm{n}}, \label{eq:massn}\\
        \frac{\partial \rho_{\rm{c}}}{\partial t} + \nabla \cdot (\rho_{\rm{c}} \vec{u}_{\rm{c}}) = S_{\rm{c}}, \label{eq:massc}
    \end{gather}
\end{subequations}
and momentum balance 
\begin{subequations}
    \begin{gather}
    \frac{\partial}{\partial t}(\rho_{\rm{n}} \vec{u}_{\rm{n}}) + \nabla \cdot (\rho_{\rm{n}} \vec{u}_{\rm{n}} \otimes \vec{u}_{\rm{n}} + \hat{p}_{\rm{n}}) = \rho_{\rm{n}} \vec{g} + \vec{R}_{\rm{n}}, \label{eq:motn}\\
     \frac{\partial}{\partial t}(\rho_{\rm{c}} \vec{u}_{\rm{c}}) + \nabla \cdot (\rho_{\rm{c}} \vec{u}_{\rm{c}} \otimes \vec{u}_{\rm{c}} + \hat{p}_{\rm{c}} + \hat{p}_{\rm{m}}) = \rho_{\rm{c}} \vec{g} + \vec{R}_{\rm{c}}, \label{eq:motc}
    \end{gather} 
\end{subequations}
for the neutrals and charges, respectively \footnote{We denote the outer product as `$\otimes$'.}. Here, $\rho_{\rm{n,c}}$ is density, $\vec{u}_{\rm{n,c}}$ the center of mass velocity and $\hat{p}_{\rm{n,c}}$ the stress tensor. Two external forces are included: gravity through the term proportional to the gravitational acceleration $\vec{g}$ and the electromagnetic force through the divergence of the magnetic stress tensor $\hat{p}_{\rm{m}}$. Energy balance equations and induction equation are not provided as they are not needed for the current discussion, but they can be found e.g., in \cite{Pop2020aa}.

The coupling between the fluids is set by the collisional terms $S_{\rm{n,c}}$ and $\vec{R}_{\rm{n,c}}$ in mass and momentum equations, respectively. The $\vec{R}_{\rm{n,c}}$ terms account for the sum of the contributions of hard sphere collisions, resonant CX, and ionization/recombination, while the $S_{\rm{n,c}}$ terms account only for ionization/recombination.

The particular expressions of the collisional terms for each kind of interaction are given later in Sections 2\ref{sec:2.1}, \ref{sec:2.2}, and \ref{sec:2.3}, respectively. The coupling terms $\vec{R}_{\rm{n,c}}$, describe the momentum transfer due to the interaction between neutrals and charges. If the components are highly coupled by collisions, the plasma moves as a whole. On another extreme, if the exchange of momentum or energy over the characteristic scales (either temporal or spatial) becomes inefficient due to the lack of coupling, then differences in the motion of the components arise. The $\vec{R}_{\rm{n,c}}$ terms can be split into elastic (if mass, momentum, and energy are conserved) or inelastic (if at least one of the former quantities is not conserved).

The components of the stress tensor for each fluid $\alpha \in \{\rm{n,c}\}$ are defined as
\begin{subequations}
   \label{eq:stresstens}
   \begin{equation}
       \hat{p}_{\rm{\alpha ij}} = p_{\rm{\alpha}}\hat{\delta}_{ij} - \frac{p_{\rm{\alpha}}}{\nu_{\rm{\alpha\alpha}}} \left(\frac{\partial u_{\rm{\alpha i}}}{\partial x_{\rm{j}}} + \frac{\partial u_{\rm{\alpha j}}}{\partial x_{\rm{i}}}\right), \tag{\ref{eq:stresstens}} 
   \end{equation}
   \begin{gather}
   p_{\rm{\alpha}} = \frac{\rho_{\rm{\alpha}}}{\tilde{\mu}_{\rm{\alpha}} m_{\rm{H}}} k_{\rm{B}} T_{\rm{\alpha}}, \label{eq:idealg}\\ 
\nu_{\rm{\alpha\beta}} = n_{\rm{\beta}} \frac{m_{\rm{\beta}}}{m_{\rm{\alpha}} + m_{\rm{\beta}}}\tilde{\sigma}_{\rm{\alpha\beta}}\sqrt{v_{\rm{T\alpha}}^2 + v_{\rm{T\beta}}^2},
   \label{eq:collfreq}
\end{gather}
\end{subequations}
where $\hat{\delta}_{\rm{ij}}$ is Kronecker's delta, $k_{\rm{B}}$ is Boltzmann's constant, $m_{\rm{H}}$ is hydrogen mass, $\tilde{\mu}_{\rm{n}} = 1$ and $\tilde{\mu}_{\rm{c}} = 0.5$ are the average mean molecular weights of neutrals and charges, $T_{\rm{\alpha}}$ is the temperature, $\nu_{\rm{\alpha\beta}}$ is the collisional frequency, $n_{\rm{\beta}}$ and $m_{\rm{\beta}}$ are the number density and mass of species $\beta$, $m_{\rm{\alpha}}$ is the mass of species $\alpha$, and $\tilde{\sigma}_{\rm{\alpha\beta}}$ is the collisional cross-section. Only self-collisions are considered in the definition of the stress tensor, so $\beta=\alpha$. The thermal speed is introduced as
\begin{equation}
    v_{\rm{T\alpha}} = \sqrt{\frac{8 k_{\rm{B}} T_{\rm{\alpha}}}{\pi m_{\rm{\alpha}}}}. \label{eq:vTi}
\end{equation}
The first term on the right-hand side of Eq. \eqref{eq:stresstens} is the pressure tensor, which is isotropic for an ideal gas. The second term is the viscous tensor. For coherency, similar to \cite{Bra1965aa} and other authors, we consider the plasma to be weakly compressible, so the term proportional to the velocity divergence is neglected. Note that neglecting compressibility in the definition of the viscous tensor makes it inappropriate for studies of acoustic waves and shocks. The viscous coefficients $\eta_{\rm{\alpha}} = p_{\rm{\alpha}} / \nu_{\rm{\alpha\alpha}}$ are taken from \cite{LeaLukLin2013aa} and \cite{PopLukKho2019aa}, assuming that frequencies of self-collision are much larger than those of collisions with other species and the gyrofrequencies. Following \cite{LeaLukLin2013aa, PopLukKho2019aa} we do not consider electron viscosity contribution into the charges viscosity. The cross-sections related to self-collisions are taken as $\tilde{\sigma}_{\rm{nn}} = 7.73\times10^{-19} \ \rm{m}^2$ and $\tilde{\sigma}_{\rm{cc}} = \frac{4}{3}\ln\Lambda \pi \frac{e^4}{(4\pi\epsilon_{0} k_{\rm{B}} T_{\rm{c}})^2}$ \cite{LeaLukLin2013aa}, where $e$ is the elementary charge and $\epsilon_{\rm{0}}$ is the vacuum permittivity\footnote{Note that a different value is stated in \cite{PopLukKho2019aa}, $\tilde{\sigma}_{\rm{nn}} = 2.1\times10^{-18}$ m$^2$, which is a typo. The same cross section as in \cite{LeaLukLin2013aa} is used in the actual computations presented here.}. The Coulomb logarithm, $\ln\Lambda$, for temperatures $T_{\rm{c}} < 50 \ \rm{eV}$ is given by 
\begin{equation}
    \ln\Lambda = 23.4 - 1.15 \log n_{\rm{e}} + 3.45 \log T_{\rm{c}},
    \label{eq:Lambda}
\end{equation}
with the electron number density  $n_{\rm{e}}=n_{\rm{c}}/2$ given in cm$^{-3}$ and $T_{\rm{c}}$ in $\rm{eV}$ \cite{Bra1965aa}. Further details about the cross-sections are presented in Appendix \ref{app:cross}.

Regarding the description of magnetic forces, we introduce the magnetic tensor as
\begin{equation}
    \hat{p}_m = \frac{B^2}{2 \mu_{\rm{0}}}\hat{I} - \frac{1}{\mu_0}\vec{B}\otimes\vec{B}, \label{eq:magtens}
\end{equation}
where $\vec{B}$ is the magnetic field vector, $\mu_{\rm{0}}$ is the vacuum magnetic permeability and $\hat{I}$ is the ($3\times3$) identity matrix.

As shown by Eqs. (\ref{eq:motn}) and (\ref{eq:motc}), both fluids are not affected by the same forces. For instance, the charged fluid directly feels the action of the magnetic field whilst the neutral fluid does not, which leads to differences in the evolution of each fluid. This kinetic decoupling can be described in terms of the difference in the charges and neutral velocities, that is, drift velocity, given by $\vec{w} = \vec{u}_{\rm{c}} - \vec{u}_{\rm{n}}$. The following sections discuss the 1F approximation to the drift velocity, comparing different contributors into its mathematical expression.

%%%%%%%%%%%%%%%%%%%%%%%%%%%%%%%%%%%%%%%%%%%%%%%%%%%%%%%%%%%%%
\subsection{Hard sphere collisions}
\label{sec:2.1}
%%%%%%%%%%%%%%%%%%%%%%%%%%%%%%%%%%%%%%%%%%%%%%%%%%%%%%%%%%%%%

Elastic collisions between plasma components (e.g., neutral-neutral, neutral-ion, neutral-electron, electron-ion, ...) are one of the dominant coupling mechanisms in the lower solar atmosphere \cite[see, e.g., Figure 1a in][]{KhoColDia2014aa}. They can be modeled as hard sphere collisions, so the cross-section is a constant that only depends on the size of the particles \cite{SchNag2009aa}. For a hydrogen plasma the momentum transfer terms due to elastic collisions are given by \cite{Bra1965aa, Dra1986aa, LeaLukLin2012aa, PopLukKho2019aa},
\begin{gather}
    \vec{R}^{\rm{elastic}}_{\rm{n}} = \alpha \rho_{\rm{c}} \rho_{\rm{n}} (\vec{u}_{\rm{c}} - \vec{u}_{\rm{n}}) = - \vec{R}^{\rm{elastic}}_{\rm{c}},
    \label{eq:Relas}
\end{gather}
with the collisional parameter
\begin{equation}
    \alpha(T) \approx \frac{1}{2m_{\rm{H}}}\sqrt{v_{\rm{Tc}}^2 + v_{\rm{Tn}}^2} \tilde{\sigma}_{\rm{pn}}, \label{eq:alpha}
\end{equation}
defined in such a way that
\begin{subequations}
    \begin{gather}
    \nu_{\rm{cn}} = \alpha \rho_{\rm{n}}, \label{eq:nucn}\\
    \nu_{\rm{nc}} = \alpha \rho_{\rm{c}},
    \label{eq:nunc}
\end{gather}
\end{subequations}
with $\tilde{\sigma}_{\rm{p n}} = 1.0 \times 10^{-18}$ m$^{2}$ being the proton-neutral cross-section taken from \cite{VraKrs2013aa}. The correctness of taking this value is discussed in Section 2\ref{sec:2.3} below. The definition of the collisional parameter given by Eq. (\ref{eq:alpha}) assumes that drift velocities are smaller than thermal speeds \cite{Dra1980aa}. It also neglects the contribution of collisions with electrons, since it is a factor of $\sqrt{m_e}$ smaller than the contribution of collisions with protons (where $m_{\rm{e}}$ denotes the electron mass). The last equality in Eq. \eqref{eq:Relas} is the consequence of the momentum conservation. 

%%%%%%%%%%%%%%%%%%%%%%%%%%%%%%%%%%%%%%%%%%%%%%%%%%%%%%%%%%%%%
\subsection{Ionization and recombination\label{sec:2.2}}
%%%%%%%%%%%%%%%%%%%%%%%%%%%%%%%%%%%%%%%%%%%%%%%%%%%%%%%%%%%%%

In this section, we focus on the ionization/recombination processes for the case of pure hydrogen plasma. Following \cite{KhoColDia2014aa}, and neglecting the momentum carried by photons, the momentum transfer terms are given by
\begin{equation}
    \vec{R}^{\rm inelastic}_{\rm{l}} = m_{\rm{H}}\sum_{\rm{k}\neq \rm{l}}^{\rm N} (n_{\rm{k}} \vec{u}_{\rm{k}}P_{\rm{kl}} - n_{\rm{l}} \vec{u}_{\rm{l}}P_{\rm{lk}}), \label{eq:Rinelas}  
\end{equation}        
where the sub-indices l,k go from 1 to the number of energy levels N, ordered from the ground level up to the most energetic one; $P_{\rm{lk}}$ is the rate of the processes that populate the level $\rm{k}$ from level $\rm{l}$, accounting for both collisional, $C_{\rm{lk}}$, and radiative ones, $R_{\rm{lk}}$ ($P_{\rm{lk}} = C_{\rm{lk}} + R_{\rm{lk}}$). For the case of the hydrogen atom considered here, protons have the highest energy level, $\rm{l} = \rm{N}$, and neutrals have levels $\rm{l} =\{1,...\rm{N}-1\}$. By assuming that all the neutrals move at the same velocity $\vec{u}_l = \vec{u}_n, \forall l \in \{1,\rm{N} - 1\}$, and neglecting the drift velocity between electrons and protons ($\vec{u}_c = \vec{u}_p = \vec{u}_e$ = $\vec{u}_{\rm N}$), the inelastic momentum exchange for the neutrals becomes,
\begin{equation}
    \vec{R}^{\rm{inelastic}}_{\rm{n}} = \sum_{\rm{l}}^{\rm{N} - 1} \vec{R}^{\rm{inelastic}}_{\rm{l}} = \rho_{\rm{c}} \vec{u}_{\rm{c}} P^{\rm{rec}} - \rho_{\rm{n}} \vec{u}_{\rm{n}} P^{\rm{ion}},
    \label{eq:Rninelas}
\end{equation}
where $\rho_{\rm{n}} = \sum_{\rm{l}}^{\rm{N} - 1} \rho_{\rm{l}}$. Formally, the recombination term must be proportional to the proton mass density, however, since protons are more massive than electrons, and there are the same number of them, $\rho_{\rm{N}}=m_{\rm{H}}n_{\rm{N}}\approx \rho_{\rm{c}}$. The rates in Eq. (\ref{eq:Rninelas}) are given by
\begin{subequations}
    \begin{gather}
    P^{\rm{rec}} = \sum_{l}^{\rm{N} - 1} P_{\rm{Nl}}, \label{eq:grec_def}\\
    P^{\rm{ion}} = \frac{1}{\rho_{\rm{n}}}\sum_{\rm{l}}^{\rm{N} - 1} \rho_{\rm{l}}P_{\rm{lN}}.
    \label{eq:gion_def}
\end{gather}
\end{subequations}

Based on the previous assumptions, the bound-bound (b-b) transitions, which take place between different excitation states within the same ionization state, are not going to lead to net momentum exchange \cite{KhoColDia2014aa}. Therefore, the recombination and ionization processes are the only contributions. All of those simplifications allow us to compact the notation in \eqref{eq:Rninelas}. Furthermore, as exchanging momentum between hydrogen charges and neutrals is the only possibility, it follows that,
\begin{equation}
\vec{R}^{\rm{inelastic}}_{\rm{n}} = - \vec{R}^{\rm{inelastic}}_{\rm{c}},
    \label{eq:mom_cons}
\end{equation}
which means that the momentum is conserved in the ionization and recombination processes. Moreover, the source terms in the mass balance equations (Eqs. \eqref{eq:massn}, \eqref{eq:massc}) are 
\begin{equation}
    S_{\rm{n}} = \rho_{\rm{c}} P^{\rm{rec}} - \rho_{\rm{n}} P^{\rm{ion}} = - S_{\rm{c}}.
    \label{eq:SE}
\end{equation}
In the case $S_{\rm{n}} = 0$, the ionization and recombination processes are in balance, leading to SE conditions. In a general situation, ionization and recombination processes are out of balance, $S_{\rm{n}} \ne 0$, leading to NEQ. The role of NEQ for the charge-neutral drift velocity is going to be discussed later.

%%%%%%%%%%%%%%%%%%%%%%%%%%%%%%%%%%%%%%%%%%%%%%%%%%%%%%%%%%%%%
\subsection{Resonant CX collisions\label{sec:2.3}}
%%%%%%%%%%%%%%%%%%%%%%%%%%%%%%%%%%%%%%%%%%%%%%%%%%%%%%%%%%%%%

Charge exchange is a process in which at least one electron is transferred from one of the colliding particles to the other. Here, we focus on the case of CX collisions between atoms of different ionization state but the same chemical species, although they can occur between different chemical species.
If the relative microscopic velocity $g_{\rm{\alpha,\beta}} = |\vec{v}_{\rm{\alpha}} - \vec{v}_{\rm{\beta}}|$ between the two colliding particles can be considered much smaller than their typical velocities ($g_{\rm{\alpha,\beta}}\ll\vec{u}_{\rm{\alpha}}, \vec{u}_{\rm{\beta}}$), the transitions between the excited states of close energy, say $\Delta E$, are the most probable \cite{Smi2003aa}. Moreover, if the particles obtained as a result of a CX collision are the same as before, as it is in the case of a pure hydrogen plasma,
\begin{equation}
    \rm{H}^{+} + \rm{H} \rightarrow \rm{H} + \rm{H}^{+},
    \label{eq:RCX}
\end{equation}
the gap of energy is $\Delta E= 0$. This kind of process is referred to as `resonant'. In general, charge exchange processes are inelastic, but in the particular case of the resonant ones, they fulfil all the conditions of the elastic collisions. 

As it is shown in Figure 1(b) of \cite{KrsSch1999aa} and Figure A3.5 of \cite{Ger2004aa}, resonant CX collisions dominate over hard sphere collisions in the charge-neutral momentum transfer of hydrogen plasmas \cite{OliSolTer2016aa}. Whilst the hard sphere collisions occur more frequently, resonant CX ones are more efficient in transferring momentum \cite{SchOvcSta2016aa}. Most of the expressions for modelling the resonant CX contribution in solar literature come from the studies of interaction of solar wind with interplanetary gas \cite{PauZanWil1995aa}, where collisions take place at very high energies ($T \gg 1\text{eV}$). This can be the case of the solar corona or the transition region, but for the chromosphere and photosphere, where the collisional energies are lower, these expressions should be applied with caution \cite{KrsSch1999aa,VraKrs2013aa}.

At low energies, the rigorous treatment of CX collisions becomes specially tricky for resonant events. In this situation, it is not possible to distinguish if the outcoming particles come from a hard sphere or a resonant CX collision due to a significative interference between these mechanisms. This way, it cannot be said, e.g., if the outcoming neutral was the incoming one or the incoming ion so, in this sense, particles become indistinguishable. Although both processes cannot be distinguished by the scattering angle or projectile energy, colliding particles can be labelled by their spin \cite{KrsSch1999aa}. In the works \cite{KrsSch1999aa, VraKrs2013aa}, the authors computed the cross-section of the elastic collisions as the coherent sum of both processes (resonant CX and hard sphere collisions), by applying the spin statistics. They found that in the case of low energies ($T \ll 1\text{eV}$), the total cross-section shows oscillations due to the quantum interference and that the magnitude of the cross-section is larger than for high energies.  In the case of high energies, the classical limit is recovered, and particles coming from the hard sphere and resonant CX collisions become distinguishable. This means that the total cross-section can be computed as the incoherent sum of both the processes, i.e., the sum of the cross-sections.
Regarding the momentum exchange rates presented in Section 2.\ref{sec:2.1}, this classical limit leads to the inclusion of an additional term in the collisional parameter $\alpha$, Eq. (\ref{eq:alpha}), which we show later.

The approach based on the distinguishable particles has been widely used in the solar 2F literature \cite[e.g.,][]{Mei2011aa, LeaLukLin2012aa, PopLukKho2019aa, SnoDruHil2023aa}. It has the advantage that both hard spheres and CX processes can be simply added by separate. However, it should be noted that its validity is restricted to the regions with higher collisional energies, as e.g., transition region and the corona. For lower layers, this approach leads to underestimation of the coupling between neutral and ionized hydrogen by a factor somewhat below 10 \cite{VraKrs2013aa}.

It has to be noted that in some works \cite[e.g.,][]{OliSolTer2016aa, PopLukKho2021aa} the value of the cross-section from \cite{VraKrs2013aa} (corrected for CX) was used as if it was the one for solely hard sphere collisions, i.e. $\tilde{\sigma}_{pn} = 1 \times 10^{-18}$ m$^2$ (see Section \ref{sec:2.1}). However, this value contains the coherent sum of both hard sphere and CX collisional cross sections. Therefore, works \cite{OliSolTer2016aa, PopLukKho2021aa} counted the CX collisions twice, overestimating the coupling. While this is not consistent, this error is within a typical range of precision of determination of collisional cross sections. To be consistent with the simulations shown in section \ref{sec:5}, we keep using $\tilde{\sigma}_{pn} = 1 \times 10^{-18}$ m$^2$ for the hard sphere collisions, keeping in mind that a more consistent option would be to use the value of $\tilde{\sigma}_{pn} = 5 \times 10^{-19}$ m$^2$ \cite{Ost1961aa,KhoCol2012aa}, where the contribution of the CX is removed from the elastic momentum transfer cross-section, and then adding the CX terms separately.

To the best of our knowledge, collisional integrals consistent with \cite{VraKrs2013aa} cross-sections are still a matter of discussion. There exist recent efforts in the context of solar physics for more rigorous calculations of the cross-sections. For example, in \cite{WarMarHan2022aa} the authors took the quadratic sum of the momentum transfer cross-sections of the hard sphere and CX collisions, but they point out that it is not fully consistent with \cite{VraKrs2013aa} calculations. 

For the purposes of this work, we are going to consider the distinguishable approach, so the hard sphere and the resonant CX collisions are going to be added as separate terms to the full elastic collisional term. Thus, instead of $\alpha$, we use $\alpha_{\rm{eff}}=\alpha+\alpha_{\rm{CX}}$ in Eq. (\ref{eq:Relas}), with the expression for $\alpha_{\rm{CX}}$ given below. In addition, we are going to use the assumption of small drifts, $w / v_{\rm{T\alpha}} \ll 1$, (which has already been applied in the definition of the hard sphere collisional parameter), so the CX collisional parameter \cite{PopLukKho2019aa} can be reduced to 
\begin{subequations}
    \label{eq:alphaCX}
    \begin{equation}
    \alpha_{\rm CX} = \frac{1}{m_{\rm{n}}}(V_{0}^{\rm{CX}} + V_{\rm{1n}}^{\rm{CX}} + V_{\rm{1p}}^{\rm{CX}})\tilde{\sigma}_{\rm{CX}}.
 \tag{\ref{eq:alphaCX}}
\end{equation}
As result, the terms $V_{0}^{\rm{CX}}$, $V_{\rm{1n}}^{\rm{CX}}$, and $V_{\rm{1p}}^{\rm{CX}}$ are functions of the thermal speeds only, given by
\begin{gather}
    V_0^{\rm{CX}} \approx \sqrt{v_{\rm{Tn}}^2 + v_{\rm{Tp}}^2}, \label{eq:VCX0_app}\\
    V_{1\rm{n}}^{\rm{CX}} \approx v_{\rm{Tn}}^2 \left(\frac{64}{\pi^2} v_{\rm{Tp}}^2 + 9 v_{\rm{Tn}}^2\right)^{-0.5}, \label{eq:VCX1n_app}\\
    V_{1\rm{p}}^{\rm{CX}} \approx v_{\rm{Tp}}^2  \left(\frac{64}{\pi^2} v_{\rm{Tn}}^2 + 9 v_{\rm{Tp}}^2\right)^{-0.5}, \label{eq:VCX1p_app}
\end{gather}
and $\tilde{\sigma}_{\rm{CX}}$ is the charge exchange cross-section, computed as
\begin{equation} 
    \tilde{\sigma}_{\rm{CX}} \approx 1.12 \times 10^{-18} - \frac{1}{2} \times 7.15 \times 10^{-20} \ln(v_{\rm{Tn}}^2 + v_{\rm{Tp}}^2),~~~[m^2]. \label{eq:sigmaCX}
\end{equation}
\end{subequations}

The fact that the parameter $\alpha_{\rm{CX}}$ presented here is independent of ${w}$ is going to be useful for the discussion in the next Section, in which we are going to obtain an expression for the 1F drift velocity. 

%%%%%%%%%%%%%%%%%%%%%%%%%%%%%%%%%%%%%%%%%%%%%%%%%%%%%%%%%%%%%
\section{Drift velocity and ambipolar diffusion coefficient}
\label{sec:3}
%%%%%%%%%%%%%%%%%%%%%%%%%%%%%%%%%%%%%%%%%%%%%%%%%%%%%%%%%%%%%

In the single-fluid (1F) formalism, the interaction between charges and neutrals is described by a series of non-ideal terms in the generalized Ohm's law. The derivation of the Ohm's law requires an expression for the drift velocity. In the 1F approach, this expression is necessarily an approximation since the individual charges and neutral velocities are not evolved. In order to derive the generalized Ohm's law here, we follow the steps of \cite{KhoColDia2014aa}. 

We first introduce the density fraction of charges, $\xi_{\rm{c}}$, and neutrals, $\xi_{\rm{n}}$, as:
\begin{equation}
    \xi_{\rm{c}} + \xi_{\rm{n}} = \frac{\rho_{\rm{c}}}{\rho} + \frac{\rho_{\rm{n}}}{\rho} = 1, \label{eq:xis}
\end{equation}
where $\rho = \rho_{\rm{c}} + \rho_{\rm{n}}$. Then we multiply \eqref{eq:motc} by $\xi_{\rm{n}}$ and \eqref{eq:motn} by $\xi_{\rm{c}}$ and compute the difference between both equations, obtaining
\begin{equation}
    \begin{gathered}
     \xi_{\rm{n}}\left[\frac{\partial}{\partial t}(\rho_{\rm{c}} \vec{u}_{\rm{c}}) + \nabla \cdot (\rho_{\rm{c}} \vec{u}_{\rm{c}} \otimes \vec{u}_{\rm{c}})\right] - \xi_{\rm{c}}\left[\frac{\partial}{\partial t}(\rho_{\rm{n}} \vec{u}_{\rm{n}}) + \nabla \cdot (\rho_{\rm{n}} \vec{u}_{\rm{n}} \otimes \vec{u}_{\rm{n}})\right] = \\
    - \xi_{\rm{n}}\nabla \cdot \hat{p}_{\rm{m}} - \vec{G} - \alpha_{\rm{eff}} \rho_{\rm{n}} \rho_{\rm{c}} \vec{w} - \rho_{\rm{c}} \vec{u}_{\rm{c}} {\rm{P}}^{\rm{rec}} + \rho_{\rm{n}} \vec{u}_{\rm{n}} \rm{P}^{\rm{ion}},
    \end{gathered}
    \label{eq:drift_whole}
\end{equation}
where $\alpha_{\rm{eff}} = \alpha + \alpha_{\rm{CX}}$ accounts for the hard sphere and resonant CX collisions and
\begin{equation}
    \vec{G} = \xi_{\rm{n}} \nabla \cdot \hat{p}_{\rm{c}} - \xi_{\rm{c}} \nabla \cdot \hat{p}_{\rm{n}},
    \label{eq:Gterm}
\end{equation}
is the thermal pressure gradient function \cite{Bra1965aa}. 

\noindent The 1F velocity of the center of mass of neutrals and charges is introduced as, 
\begin{equation}
    \vec{u} = \frac{\rho_{\rm{n}} \vec{u}_{\rm{n}} + \rho_{\rm{c}} \vec{u}_{\rm{c}}}{\rho_{\rm{n}} + \rho_{\rm{c}}}. \label{eq:u}
\end{equation}

\noindent Given this definition, we can introduce the drift velocity of the neutral and charged species with respect to the center of mass velocity of the whole fluid,
\begin{subequations}
    \begin{gather}
     \vec{u}_{\rm{n}} - \vec{u} = \vec{u}_{\rm{n}} - \frac{\rho_{\rm{n}} \vec{u}_{\rm{n}} + \rho_{\rm{c}}
    \vec{u}_{\rm{c}}}{\rho_{\rm{n}} + \rho_{\rm{c}}} = - \xi_{\rm{c}} \vec{w},  \label{eq:wn} \\
     \vec{u}_{\rm{c}} - \vec{u} = \vec{u}_{\rm{c}} - \frac{\rho_{\rm{n}} \vec{u}_{\rm{n}} + \rho_{\rm{c}}
    \vec{u}_{\rm{c}}}{\rho_{\rm{n}} + \rho_{\rm{c}}} = \xi_{\rm{n}} \vec{w},    \label{eq:wc}
\end{gather}
\end{subequations}
and use them to substitute individual velocities, $\vec{u}_n$ and $\vec{u}_c$, in favor of $\vec{w}$ and $\vec{u}$ in Eq. (\ref{eq:drift_whole}).

\noindent By neglecting the left hand side terms in Eq. (\ref{eq:drift_whole}), which is reasonable for processes slower than the collisional time scales, see \cite{KhoColDia2014aa,DiaKhoCol2014aa}, we obtain an approximate expression for the drift velocity
\begin{gather}
    \vec{w}_{\rm{1F}} = \tilde{\eta}_{\rm{A}} [- \xi_{\rm{n}} \nabla \cdot \hat{p}_{\rm{m}} - \vec{G} - S_{\rm{n}} \vec{u}], \label{eq:drift_eq}
\end{gather}
where a small term proportional to the current density, $\vec{J}$, multiplied by electron mass, $m_{\rm{e}}$, has not been included, compared to the equation (133) in \cite{KhoColDia2014aa}. To simplify the discussion below, we labeled this approximate drift velocity as $\vec{w}_{\rm{1F}}$. The coefficient in front of the square brackets in Eq. (\ref{eq:drift_eq}), $\tilde{\eta}_{\rm{A}}$, is proportional to the ambipolar diffusion coefficient, $\eta_A=\tilde{\eta}_{\rm{A}}\xi_n^2B^2$, and is equal to,
\begin{equation}
\tilde{\eta}_{\rm{A}} = \frac{1}{\alpha_{\rm{eff}} \rho_{\rm{n}} \rho_{\rm{c}}  + \rho_{\rm{c}} \xi_{\rm{n}} \rm{P}^{\rm{rec}} + \rho_{\rm{n}} \xi_{\rm{c}} \rm{P}^{\rm{ion}}}, \label{eq:etaA}
\end{equation}

\noindent There are two modifications in comparison with the usual expressions for the drift velocity and ambipolar diffusion coefficient, namely

\begin{itemize}
    \item The coefficient $\tilde{\eta}_A$ depends on the elastic collisions as well as on the inelastic ones.
    \item Apart from the usual magnetic drift (mainly referred to as the ambipolar one) and the thermal pressure gradient function drift \cite{KhoColDia2014aa, BalAleCol2018aa}, there is a new drift component due to the inelastic collisions if the plasma is out of statistical equilibrium (i.e. if $S_n \ne 0$). 
\end{itemize}

\noindent In order to evaluate the importance of the inelastic component into the ambipolar coefficient, it is useful to introduce the ratio
\begin{equation}
    \varepsilon = \frac{\rho_{\rm{c}} \xi_{\rm{n}} \rm{P}^{\rm{rec}} + \rho_{\rm{n}} \xi_{\rm{c}} \rm{P}^{\rm{ion}}}{\alpha_{\rm{eff}} \rho_{\rm{n}} \rho_{\rm{c}}}.
    \label{eq:ineltoelas}
\end{equation}

The drift velocity $\vec{w}_{\rm{1F}}$ defined in Eq. (\ref{eq:drift_eq}), depends on the vector $\vec{G}$, which is a function of the stress tensors (Eq. \ref{eq:stresstens}) and, therefore, of the derivatives of the individual charges and neutral velocities, if viscosity is considered. Thus, further approximations are needed in order to consistently use Eq. (\ref{eq:drift_eq}) in the 1F case. A reasonable assumption is to consider that $w \ll u_{\rm{n,c}}$, so $\frac{\partial u_{\rm{\alpha i}}}{\partial x_{\rm{j}}} \approx \frac{\partial u_{\rm{i}}}{\partial x_{\rm{j}}}$. This way, the drift velocity can be computed from Eq. (\ref{eq:drift_eq}) if viscosity is taken into account.  
 
In the sections below, we compare the contribution of elastic and inelastic effects, and of the pressure gradient function onto the drift velocity under different conditions.

%%%%%%%%%%%%%%%%%%%%%%%%%%%%%%%%%%%%%%%%%%%%%%%%%%%%%%%%%%%%%
\section{Comparison under SE conditions}
\label{sec:4}
%%%%%%%%%%%%%%%%%%%%%%%%%%%%%%%%%%%%%%%%%%%%%%%%%%%%%%%%%%%%%

In this section, we consider the VAL3C 1D model atmosphere of the quiet Sun \cite{VerAvrLoe1981}.
It spans from the photosphere to the base of the corona. From the model we adopt the height profiles of temperature, $T$,
the total number density of hydrogen, $n_{\rm{H}}$, and the electron number density $n_{\rm{e}}$.

\begin{figure}[t]
\centering
\includegraphics[width=0.8\hsize]{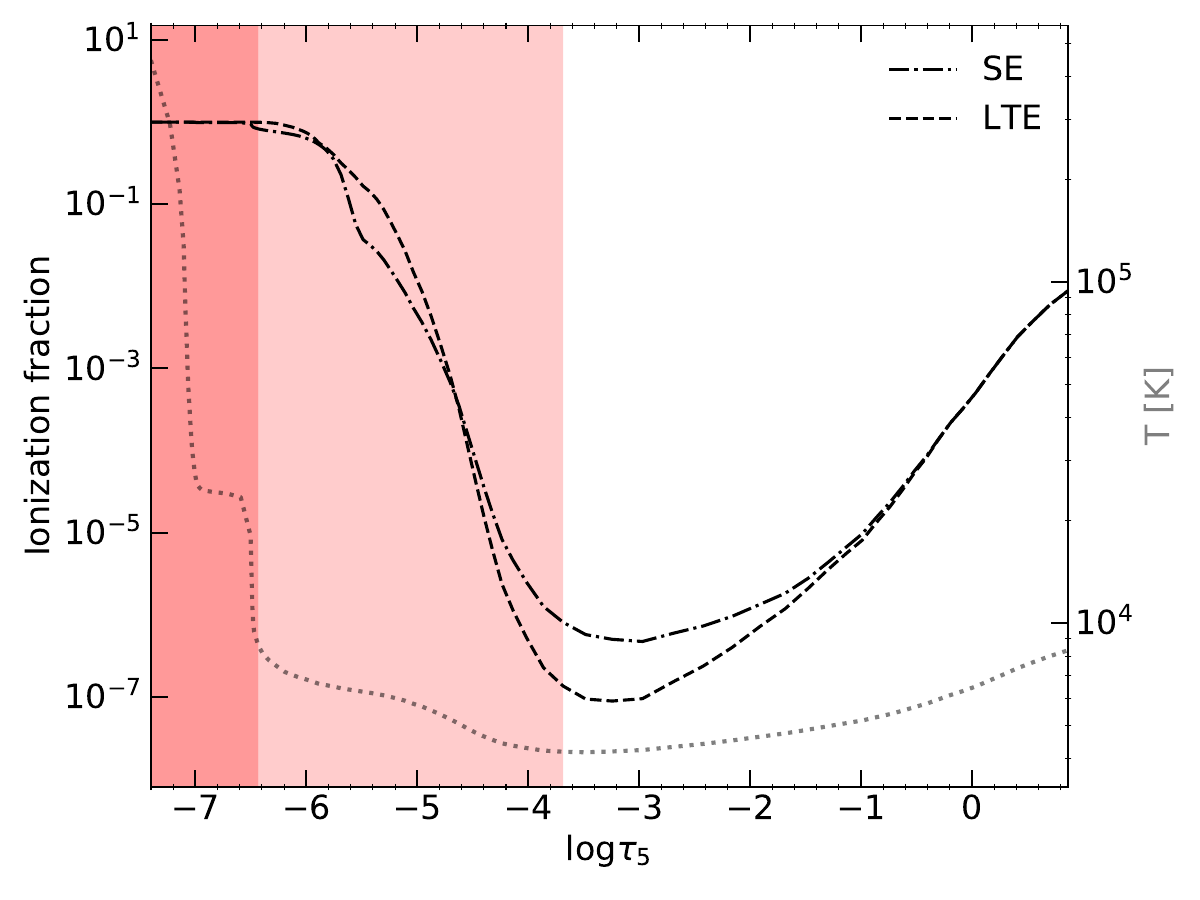}
\caption{Temperature profiles of the VAL3C model (dotted line), and hydrogen ionization fractions in LTE (dashed line) and in SE (dashed-dotted line) considering a (5 + 1 levels) hydrogen model as functions of logarithm of the optical depth $\tau_{5}$ at 500 nm. The filled regions, from left to right, represent approximately the extension in optical depth of the different layers of the Sun: transition region and coronal base, chromosphere, and photosphere.}
\label{fig:Ifrac}
\end{figure}

Figure \ref{fig:Ifrac} shows the temperature and the hydrogen ionization fraction ($\xi_{\rm{c}} = n_{\rm{i}} / (n_{\rm{n}} + n_{\rm{i}})$) of the model atmosphere as functions of the logarithm of the continuum optical depth at 500 nm, $\tau_5$.
The ionization fraction is computed using the MULTI code \cite{Car1986aa}, which solves the radiative transfer equation (RTE) together with the rate equations (Eq. \ref{eq:massn} and \ref{eq:massc}, but in terms of $n_{\rm{\alpha}}$) assuming statistical equilibrium (SE), for a given model atom. Only hydrogen populations are computed to be consistent with the radiation field, whereas the other chemical species are just contributing to the opacity. The code provides populations for each atomic level, in addition to the corresponding radiative and collisional transition rates \cite{Rut2003aa}. For the present work, we consider a hydrogen model atom of 5 + 1 levels, with 5 bound levels and one ionized state, i.e., protons. We then computed hydrogen populations using either the solution from MULTI (i.e. SE for the rate equations and NLTE conditions for the radiation field), or the populations from the first iteration step, corresponding to the LTE conditions for the radiation field. We will refer to the first computation as SE, and to the second one as LTE. As expected, under given conditions, Fig. \ref{fig:Ifrac} shows that hydrogen is almost fully ionized in the region of $\log{\tau_{5}} < -6$ and temperatures of $T \gtrsim 6 \times 10^{3} \ \rm{K}$; then, the ionization fraction drastically decreases in the intermediate region of $\log{\tau_{5}}$ and low temperatures, and finally starts to increase again for larger values of the optical depth. SE case (dashed-dotted line in Fig. \ref{fig:Ifrac}) gives larger hydrogen ionization fractions around the temperature minimum, and lower ones higher in the chromosphere, compared to the LTE case (dashed line). 

%NV From this point onward, we ignore any other specie in the model atmosphere but hydrogen, and we continue our analysis as if the VAL3C model was constructed for pure hydrogen atmosphere. While this assumption is obviously not accurate, it provides us with realistic temperature structure and realistic distribution of the hydrogen number densities, although they are not strictly consistent with physics of the VAL3C model. 

\begin{figure}[t]
\centering
\includegraphics[width=0.7\hsize]{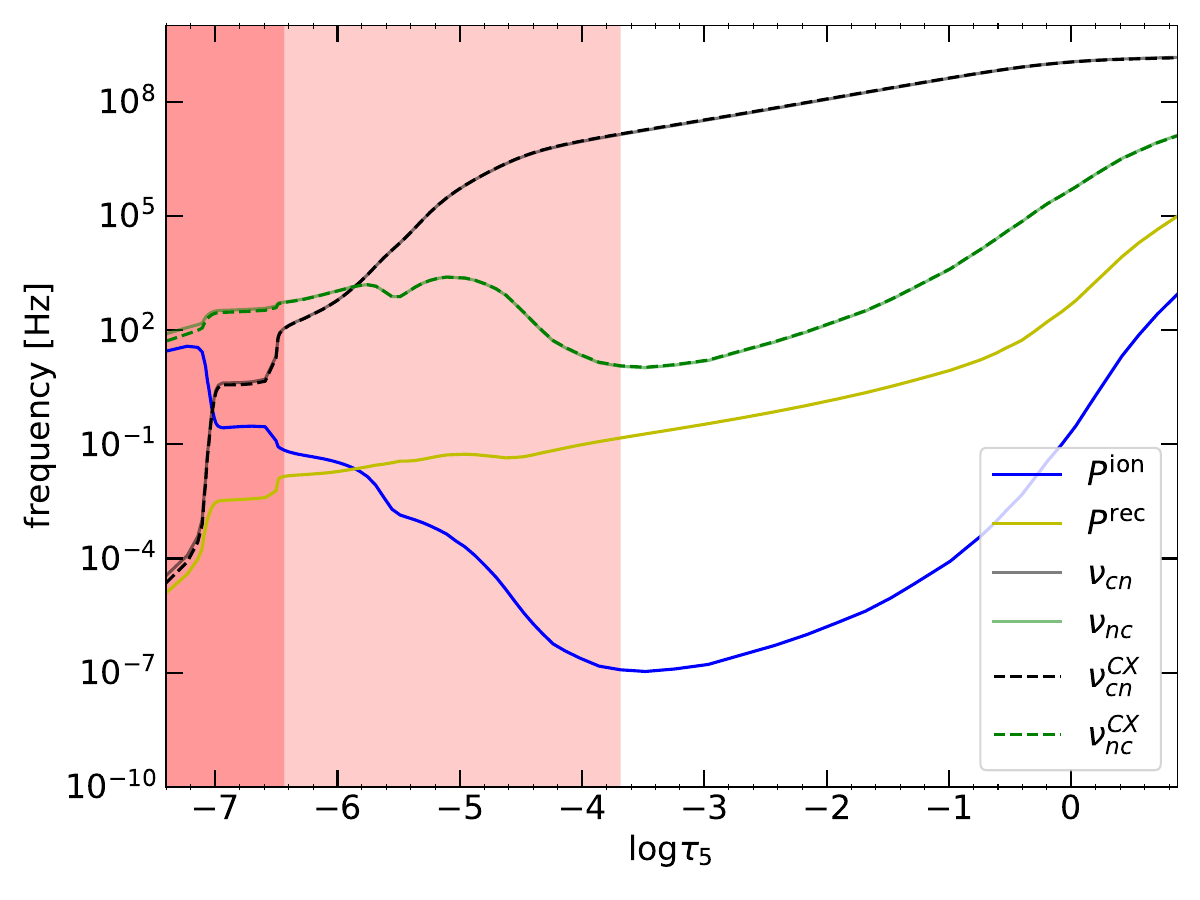}
\caption{Elastic and inelastic collisional frequencies for the solar atmospheric model from Fig. \ref{fig:Ifrac} as functions of $\tau_{5}$. Blue and yellow lines, respectively, represent the ionization and recombination rates (Eq. \eqref{eq:gion_def} and \eqref{eq:grec_def}). Green and black lines, respectively, are for the elastic neutral-charge and charge-neutral collisional frequencies. Dashed lines are for the hard sphere collisions, while dashed-dotted lines are for the CX collisions.}
\label{fig:IR}
\end{figure}

\begin{figure}[t]
\centering
\includegraphics[width=0.7\hsize]{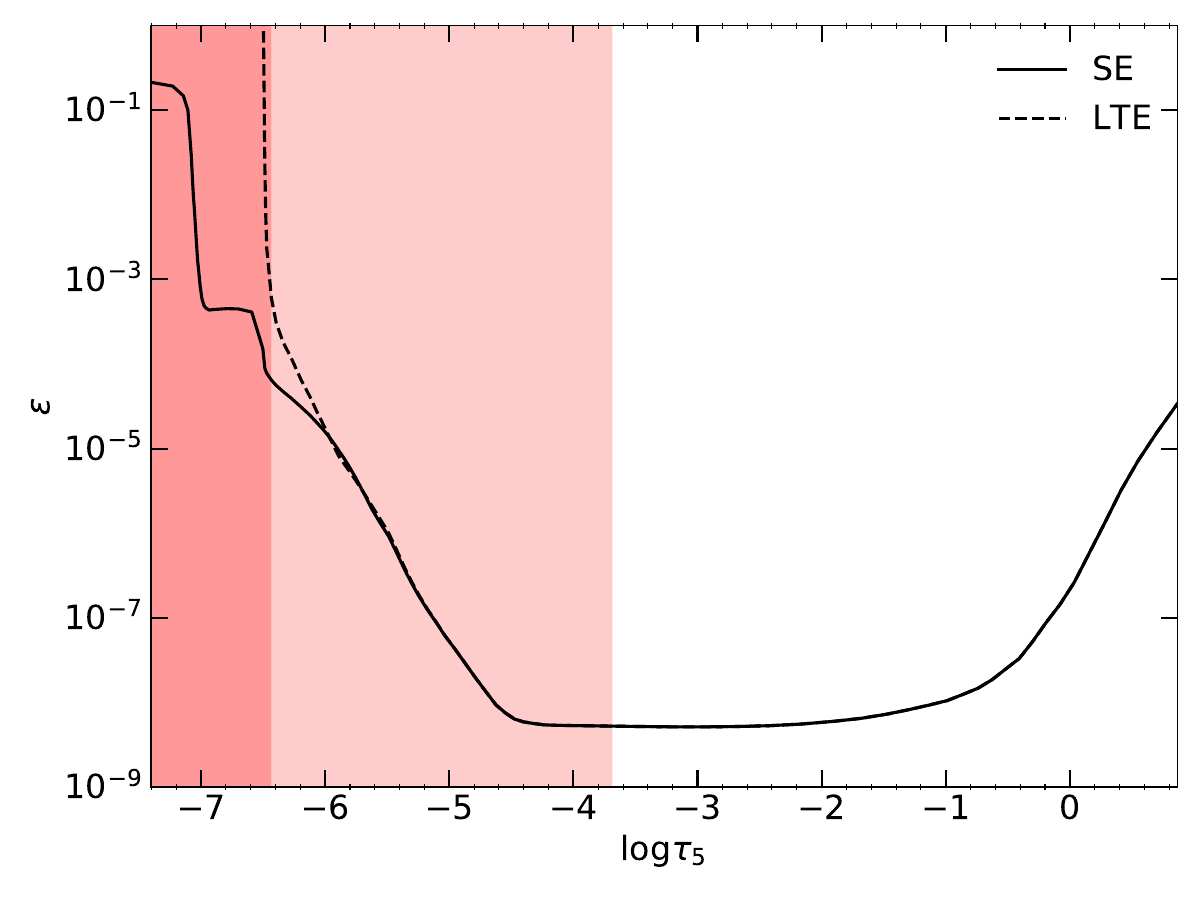}
\caption{Ratio of inelastic to elastic processes, $\varepsilon$ from Eq. (\ref{eq:ineltoelas}), as a function of $\tau_{5}$, for the solar atmospheric model from Fig. \ref{fig:Ifrac}. Dashed line: ionization and recombination rates are computed in LTE; solid line: rates are computed in SE.}
\label{fig:Ambicorr}
\end{figure}

In SE conditions, the collisional terms of the continuity equations, Eqs. (\ref{eq:massn}) and (\ref{eq:massc}), are $S_{\rm{n}} = S_{\rm{c}} = 0$. Consequently, the term of the drift velocity related to the inelastic collisions, that is, the last term of Eq. (\ref{eq:drift_eq}), is set to zero. However, the contribution of inelastic collisions to $\tilde{\eta}_{\rm{A}}$, Eq.~\ref{eq:etaA}, remains. In order to estimate the magnitude of the correction with respect to the usual ambipolar coefficient, we compare in Figure \ref{fig:IR} the momentum transfer frequencies of the different processes (elastic and CX collisions, ionization and recombination) as functions of $\log \tau_5$. Alongside, Figure \ref{fig:Ambicorr} shows the ratio of the inelastic to the elastic terms defined by Eq. (\ref{eq:ineltoelas}).

According to Fig. \ref{fig:IR}, the charge-neutral collision frequencies, $\nu_{\rm{cn}}$ and $\nu_{\rm{cn}}^{\rm{CX}}$, and the recombination rate, $P^{\rm{rec}}$, decrease as the optical depth decreases, which can be related to the drop of total plasma density as we move up in the atmosphere. In the case of elastic collisions, the frequencies are proportional both to the densities and to the square root of the temperatures, as shown by Eq. \eqref{eq:collfreq}. Since the temperature increases with height at the upper layers of this model atmosphere,  the observed behaviour of the collision frequencies is explained by the exponential decrease of density. In the case of the recombination rate, it is not straightforward to talk about the role of the temperature as the radiation is not in equilibrium with the plasma. Nevertheless, in this case, the rates are also proportional to the density. The neutral-charge collision frequencies, $\nu_{\rm{cn}}$ and $\nu_{\rm{cn}}^{CX}$, follow a similar behaviour, although they exhibit respective minima around $\log \tau_{5} \sim -3.5$, caused by the extremely low ionization fraction that is present around the minimum of temperature (as shown in Fig. \ref{fig:Ifrac}). The ionization rate has its highest values at the top of the atmosphere, where the temperature is higher, and at the bottom, for large densities. In contrast, it becomes negligible in comparison with the other frequencies as the temperature drops. Moreover, although the profiles for hard sphere and charge exchange collisions seem to overlap, a closer inspection reveals that for the former processes, the frequency is larger. Apparently, this is not in agreement with the discussion in Section 2\ref{sec:2.3}, where it was stated that resonant CX collisions dominate over hard sphere interactions in hydrogen plasmas. The reason for this discrepancy is that we computed the hard spheres and resonant CX collisions following \cite{PopLukKho2021aa, PopLukKho2021ab} which, as we discussed, is not the most appropriate approximation. Also, it should be noted that while hard sphere cross-section is constant, the one for resonant CX decreases with temperature. Regarding the ionization and recombination processes, it can be seen that the recombination rate is larger for most of the range of $\log \tau_{5}$, but it is orders of magnitude smaller than the ionization rate in the upper layers of the atmosphere.
Finally, the comparison of the frequencies of elastic (including CX) and inelastic collisions shows that, except for the upper transition region, elastic collisions dominate the coupling between the species.

In Figure \ref{fig:Ambicorr} we show the inelastic to elastic ratio $\varepsilon$ computed assuming either the LTE or SE populations of hydrogen. The two functions are nearly identical up to the chromosphere where they split as the LTE one is governed by the local temperature rise. Nevertheless, in both cases $\varepsilon \ll 1$ for almost the entire range of optical depths studied here. Only for values of $\log \tau_{5}$ smaller than $-6.5$ in the LTE case or $-7.2$ in the SE case, the contribution of inelastic collisions produces a ratio $\varepsilon > 0.1$. This leads to the conclusion that the role of the inelastic collisions in modifying the ambipolar diffusion coefficient is rather small. Consequently, the classical definition of the ambipolar coefficient can be safely used over the photosphere and the chromosphere of the Sun.

%%%%%%%%%%%%%%%%%%%%%%%%%%%%%%%%%%%%%%%%%%%%%%%%%%%%%%%%%%%%%
\section{Comparison under coronal NEQ conditions}
\label{sec:5}
%%%%%%%%%%%%%%%%%%%%%%%%%%%%%%%%%%%%%%%%%%%%%%%%%%%%%%%%%%%%%

The analysis above gives an estimation of the impact of the inelastic collisions onto the ambipolar diffusion coefficient. However, it does not allow us to evaluate the component of the drift velocity that appears under NEQ conditions, given by the last term of Eq. (\ref{eq:drift_eq}), since the analysis was performed assuming SE, when $S_n=0$. In order to assess its influence on the drift velocity, we used 2.5D two-fluid simulations of Rayleigh-Taylor Instability (RTI) computed by \cite{PopLukKho2021aa, PopLukKho2021ab, LukKhoPop2024aa}. We analyzed two simulations with the parameters identical to those listed in Table 1 of \cite{PopLukKho2021ab,PopLukKho2021aa} with the labels P and L2 (these labels are related to the configuration of magnetic field, as explained below), but with 4 times more spatial resolution, as in \cite{PopLukKho2023aa,LukKhoPop2024aa}.

A solar prominence thread was simulated in a 2.5D configuration with a magnetic field either perpendicular to the perturbation plane (for the P case) or sheared with respect to this plane by 1 degree over a distance of $L_s=L_0/2$, where $L_0=1$ Mm \cite{PopLukKho2021aa}. In the first case, the development of the instability is essentially unaffected by the magnetic forces, whilst in the second case, there is a component of the magnetic field in the perturbation plane affecting the RTI cutoff, the development of the structures and the force balance. The initial configuration, adopted from \cite{LeaDeVTha2014aa}, assumes a smooth transition between a partially ionized prominence thread, with temperature about $10^4$ K and ionization fraction $\xi_c\approx 0.09$, and almost fully ionized corona, with temperature about $2\times10^5$ K. The simulations took into account hard sphere, resonant CX and ionization/recombination collisions between hydrogen neutrals and charges, viscosity, and neutral thermal conduction. Only collisions and viscosity are discussed here because they play a role in the 1F momentum equation given by Eq. (\ref{eq:drift_whole}). It is important to mention for the following discussions that the model for viscosity applied by \cite{PopLukKho2021aa, PopLukKho2021ab, LukKhoPop2024aa} taken from \cite{Bra1965aa} has been derived assuming strongly collisionally coupled plasma and the viscosity coefficient (given by Eq. (A9) in \cite{PopLukKho2019aa}) diverges in corona due to its low density \cite{HunPasKho2022aa}. In order to avoid unrealistic large values, the maximum value of the dynamic viscosity coefficient of the neutral fluid was limited to $\max(\eta_{\rm{n}} / \rho_{\rm{n}}) = 6.25\times10^7$ m$^{2}$ s$^{-1}$. The ionization/recombination rates are computed in the coronal approximation, following \cite{MeiShu2012aa}, assuming that ionization is produced by electron impact and recombination is spontaneous. Initial conditions are set to be close to the magnetohydrostatic equilibrium and SE is not satisfied ($S_n\ne 0$), so NEQ effects will be present from the beginning. With the development of the instability, \cite{PopLukKho2023aa} noted that the ionization imbalance in the course of RTI leads to the creation of a layer of an overionized plasma surrounding the drops and spikes.

In the non-linear phase, when bubbles and spikes appear \cite{PopLukKho2021aa}, the RTI leads to the development of a turbulent flow with abundant complex small-scale structures. For the present study, we have selected a snapshot corresponding to a stationary stage in the non-linear regime of the instability. Using this snapshot we computed, on the one hand, the drift velocity by directly subtracting the charge and neutral velocities from the simulations, and, on the other hand, the contributions to the drift velocity by the different forces according to Eq. (\ref{eq:drift_eq}). Horizontal and vertical components of the drift velocity turned out to show a similar range of values. Here we focus our analysis on the vertical component, which allows us to assess the role of gravitational stratification.

\begin{figure*}[t]
\includegraphics[width=\hsize]{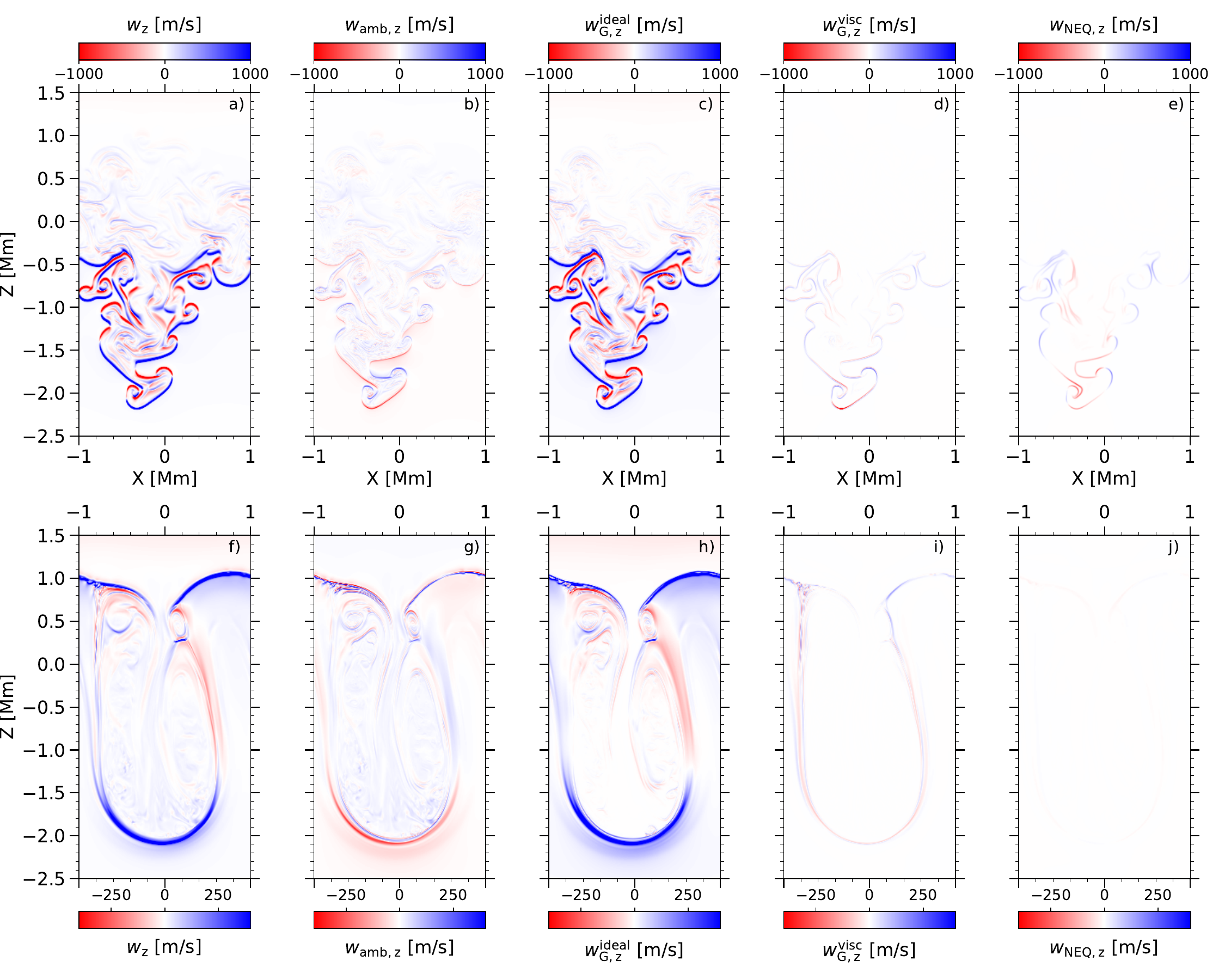}
\caption{Colour maps of the vertical drift velocity and its components according to Eq. (\ref{eq:drift_eq}) for the RTI simulation P (top row) and L2 (bottom row). From left to right: the total vertical drift velocity (computed as $w_{z} = u_{\rm{c},z} - u_{\rm{n},z}$), and the components related to ambipolar diffusion ($w_{\rm{amb},z}$), pressure gradients ($w_{\rm{G},z}^{\rm{ideal}}$), viscosity ($w_{\rm{G},z}^{\rm{visc}}$), and NEQ ($w_{\rm{NEQ},z}$), respectively.}
\label{fig:Driftmapz}
\end{figure*}

Figure \ref{fig:Driftmapz} shows colour maps of the ``true'' vertical drift velocity from the simulations together with its contributions:  $\vec{w}_{\rm{amb}}$ is the drift associated with the magnetic field, i.e., the drift due to the ambipolar diffusion, $\vec{w}^{\rm{ideal}}_{\rm{G}}$ is the drift due to the ideal part of the thermal pressure function, $\vec{w}^{\rm{visc}}_{\rm{G}}$ is the drift due to the viscous part, and $\vec{w}_{\rm{NEQ}}$ is the one due to the ionization/recombination in NEQ conditions. Panels a) and f) of Fig. \ref{fig:Driftmapz} show that the largest drifts are located at the contours of the structures, with values reaching $\pm 3$ km s$^{-1}$ in the P case, and $\pm 2$ km s$^{-2}$ in the L2 case (note that, for visualization purposes, the colour map is saturated at lower values, $\pm 1$ and $\pm 0.5$ km s$^{-1}$, respectively). Typical velocities of both fluids are of the order of $10^1$  km s$^{-1}$ in both cases \cite{PopLukKho2021ab} so, the drifts are of the order of $\sim 5-10 \%$ of the flow velocities. Therefore, the coupling is not so strong in those regions. 

\begin{figure}[t]
\centering
\includegraphics[width=0.9\hsize]{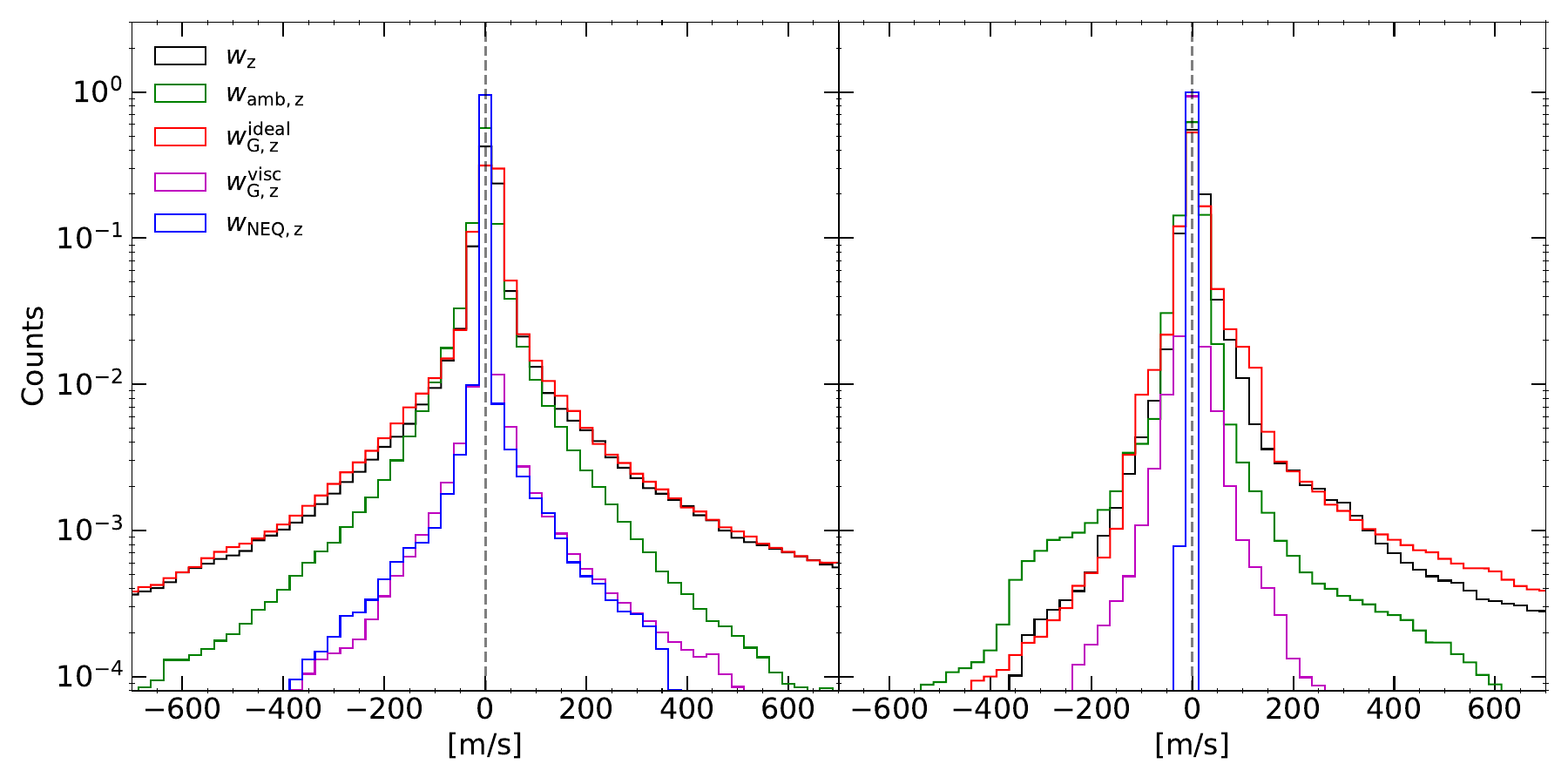}
\caption{Histogram of the contributions to $w_z$. Blue line: contribution of the NEQ term; green line: contribution from the ambipolar term; red lines: contribution from the scalar partial pressure term; purple line: contribution from the viscous one. Black line shows the drift velocity computed directly from the simulations ($w_{z} = u_{z,\rm{c}} - u_{z,\rm{n}}$). Results for the RTI P and L2 simulations are shown on the left and on the right panels, respectively.
}
\label{fig:Histz}
\end{figure}

The comparison of the panels of Fig. \ref{fig:Driftmapz} shows that in both simulations the drift velocity is dominated by the thermal pressure contribution. In particular, the ideal part of $\vec{w}^{\rm{ideal}}_{\rm{G}}$ (that is, the part related to the scalar pressure), shown at at panels c) and h), is practically indistinguishable from the ``true'' drift velocity map. Depending on the locations, contributions by magnetic pressure or viscous forces gain importance in the L2 case, while their relative contribution remains small in the P case. The histograms in Fig. \ref{fig:Histz} point in the same direction. They show that for small drift values, both ambipolar and thermal pressure gradient term contributions are similar. However, the thermal pressure term, and not the ambipolar term, is responsible for producing the largest values of the drifts manifested as tails of the histogram.

Perhaps surprisingly, we found the ambipolar contribution to play a secondary role in the creation of the drifts. In many cases in the literature, the drift velocity is approximated by the ambipolar term, i.e., magnetic forces produce the difference in motion between charges and neutrals, accounting successfully for heating and decoupling effects \cite[see, e.g.][]{MarDePHan2012aa, KhoCol2012aa,  BalAleCol2018aa, PopKep2021aa}. This approach has been justified for the case of magneto-convection simulations, or for a semi-empirical sunspot model atmosphere till chromospheric heights in \cite{KhoColDia2014aa}, by a direct evaluation of the thermal pressure function, see their Figures 8, 9. Here we show that in the conditions of the RTI, such an approach has to be taken with caution. Noting that the RTI is mainly a hydrodynamic process, it could be expected that the contributions related to the hydrodynamic pressures are going to dominate the drifts. Due to the particular conditions in the current experiment, the out-of-plane magnetic field configuration removes (in the P case) or decreases (in the L2 case) the influence of the magnetic forces compared to the hydrodynamic ones. While this situation is rather peculiar, it nevertheless could be representative for flows perpendicular to the magnetic field caused either by RTI or other processes in a 3D volume.

Whilst ambipolar and thermal pressure contributions exhibit faint substructures inside the falling drops, the viscous and NEQ contributions are significantly more spatially localized in the transition layer surrounding the drops, where the velocity shear reaches its largest values. Mostly, they play a secondary role in the drifts, but in the case of L2, Figure \ref{fig:Driftmapz} shows a region around $X = [-1, -0.6]$ Mm and $Z = [0, 1]$ Mm (see panels  f) and i)) where the viscosity term dominates. This behaviour is also evident from Fig. \ref{fig:Histz}, where we observe that the viscous term histogram is rather narrow and centered around zero. Despite the wings showing larger values, their overall contribution is small, and the decoupling due to viscosity is almost negligible over the domain. 

We obtain that the contribution of the NEQ term to the total drift velocity is the least relevant, see panels e) and j) of Fig. \ref{fig:Driftmapz}. Similar to the viscous term, it only acts over a narrow distance at the transitions between the drops and the coronal material, but in this case, the values only reach $\sim10\%$ of the maximum drift velocity. Our analysis shows that RTI does not lead to a large momentum exchange due to inelastic collisions, possibly because the flows are subsonic, with no abrupt temporal variations as it would be in shocks. This conclusion could be sensitive to the initial configuration, for example regarding how far it is from the SE, or if the evolution brings it closer or further from the SE. It is also sensitive to the model adopted for accounting for the inelastic collisions. For example, radiation and the internal structure of the hydrogen atom cannot be ignored in the solar atmosphere, therefore using a complete hydrogen atom (as in Section \ref{sec:4}), or a simplified model as in \cite{MeiShu2012aa}, would lead to different results. Nevertheless, without performing proper modeling, it is difficult to  estimate the impact. Recently, some efforts were made in this sense, as the work of \cite{SnoDruHil2023aa}, where they used a 5 + 1 level hydrogen model presented by \cite{Sol1999aa} to study the propagation of shock waves in chromospheric conditions by use of a 2F model. Even though they assume a black body radiation background and LTE conditions, it represents a necessary step to move further with more realistic physical conditions.

\begin{figure}[t]
\centering
\includegraphics[width=0.6\hsize]{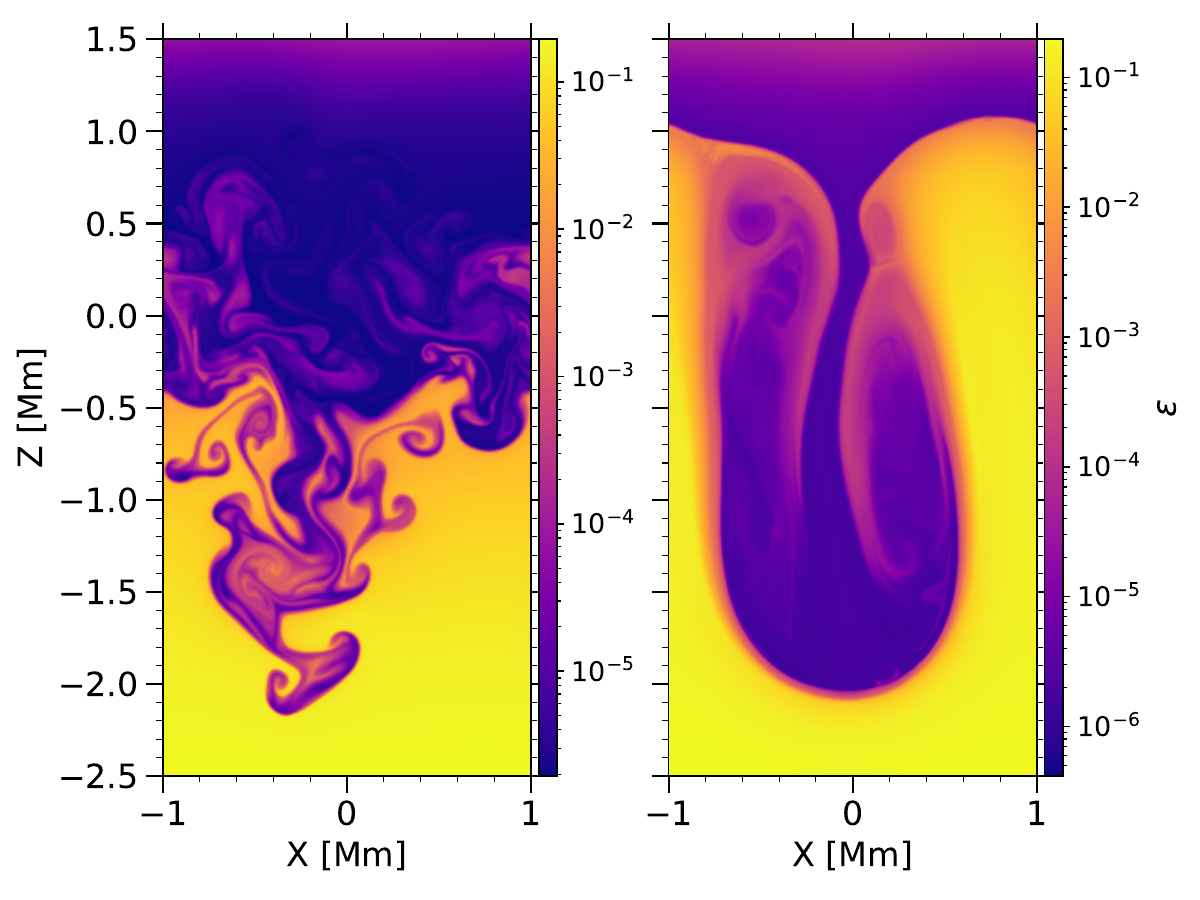}
\caption{Colour maps of the ratio of inelastic to elastic processes, $\varepsilon$, from Eq. (\ref{eq:ineltoelas}), for the RTI simulations. Left panel: P case; right panel: L2 case.}
\label{fig:Ambicorr_rt}
\end{figure}

Figure \ref{fig:Ambicorr_rt} shows the ratio of the inelastic to elastic contribution to the ambipolar coefficient, Eq. (\ref{eq:etaA}). This figure reveals two different regimes: inside the cooler prominence plasma elastic collisions are the main contribution; outside it, for the hotter plasma, the contribution from inelastic collisions reaches some $10\%$. Overall, elastic collisions seem to be  sufficient to provide a precise enough value of the ambipolar diffusion coefficient in this RTI experiment. 

\begin{figure}[t]
\centering
\includegraphics[width=\hsize]{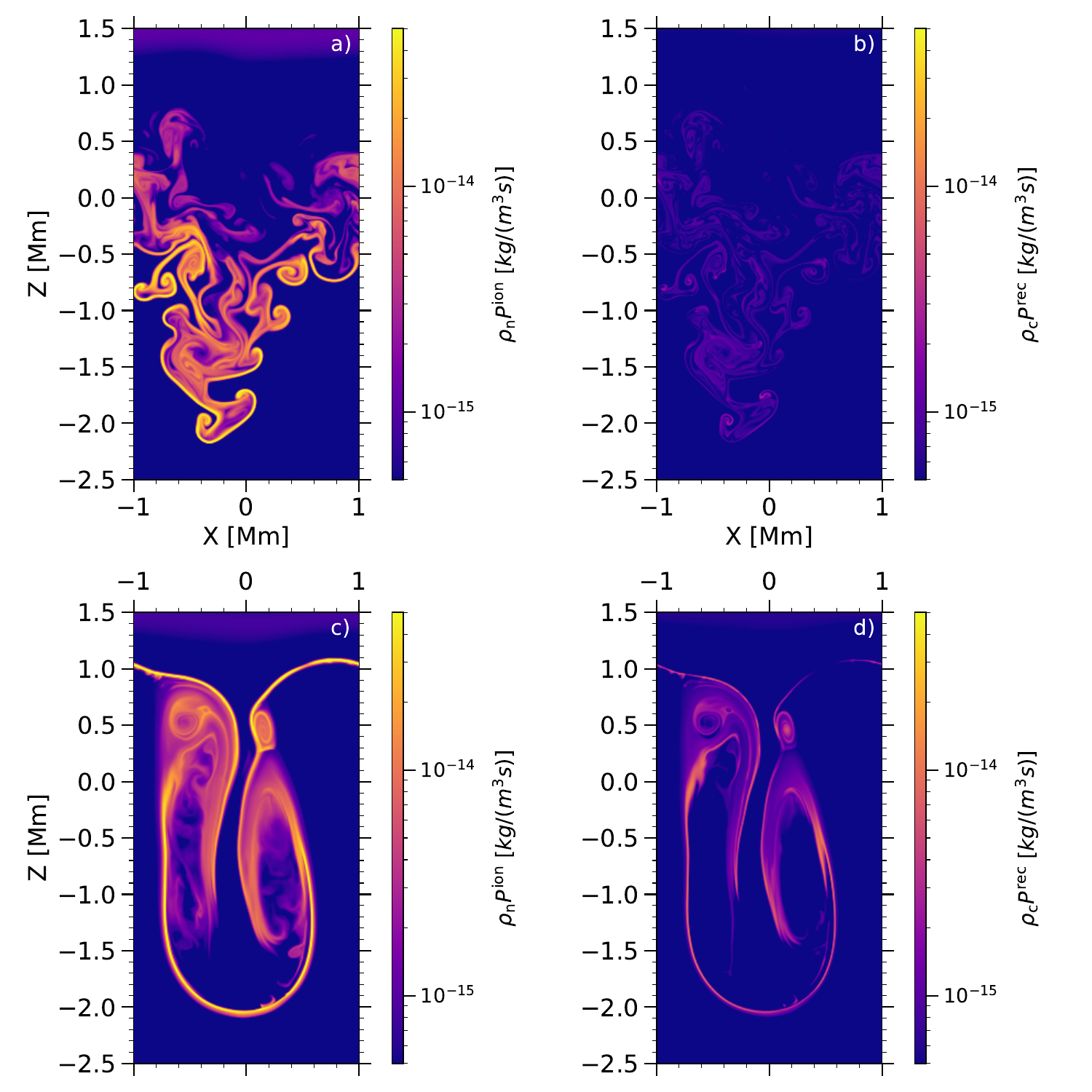}
\caption{Colour maps of the collisonal terms for mass balance ($S_{\rm{n}}$), for ionization (panels a) and c)) and recombination (panels b) and d)). The couple of panels at the top are from P simulation. Those at the bottom, for L2 case.}
\label{fig:IonRec2}
\end{figure}

To deepen on the role of the NEQ, we represent in Figure \ref{fig:IonRec2} the ionization and recombination contributions to the mass source term $S_{\rm{n}}$ (see Eq. \eqref{eq:SE}). We find that there is an excess of ionization over recombination in most of the areas of the prominence, in both simulations. One can evaluate the importance of the ionization/recombination terms in the dynamics by taking some reference values of the densities from the simulations. In coronal regions, the lowest values of density are of the order of $10^{-13}$ kg $\rm{m}^{-3}$, while in the prominence regions, the densities are approximately 1-2 orders of magnitude larger \cite{PopLukKho2021aa}. By taking the reference values for ionization/recombination to be of the order of $10^{-15}$ kg $\rm{m}^{-3} \rm{s}^{-1}$ in coronal regions and  $10^{-14}$ kg $\rm{m}^{-3} \rm{s}^{-1}$ inside the prominence, one can roughly estimate that we need about 100 s for producing a number of neutrals and charges similar to those in the regions of the lowest density. This estimate can explain the low contribution of inelastic collisions to the kinetic decoupling, as only a few particles are produced through these mechanisms. Therefore, large velocities are needed to bring or remove enough momentum from the ionizing or recombining fluid to have a significant drift velocity. It is interesting to note the fact that while ionization rates are of the same order of magnitude for P and L2 cases, the recombination rates in the P case are smaller than in the L2 case. This behaviour is a response to the different thermodynamic conditions (temperatures, densities) created in the plasma mixture in these simulations, as a consequence of their different magnetic configurations.

%%%%%%%%%%%%%%%%%%%%%%%%%%%%%%%%%%%%%%%%%%%%%%%%%%%%%%%%%%%%%
\subsection{Role of thermal pressure function in magneto-hydrostatic equilibrium} 
\label{sec:5.1}
%%%%%%%%%%%%%%%%%%%%%%%%%%%%%%%%%%%%%%%%%%%%%%%%%%%%%%%%%%%%% 

The results above show the importance of the thermal pressure function in creating the kinetic decoupling in the turbulent flow developed by the RTI. It is also interesting to verify the importance of this function for the initial equilibrium.

If the initial velocities of both the neutral and charged fluids are set to $\vec{u}_{\rm{n,0}} = \vec{u}_{\rm{c,0}} = \vec{0}$, the momentum equations of the 2F model, Eqs. (\ref{eq:motn}) and (\ref{eq:motc}), lead to the following relations:
\begin{subequations}
    \begin{gather}
    \nabla \cdot \hat{p}_{\rm{n},0}=  \rho_{\rm{n},0} \vec{g}, \label{eq:eq2F1n}\\
    \nabla \cdot(\hat{p}_{\rm{c},0} + \hat{p}_{\rm{m},0}) = \rho_{\rm{c},0} \vec{g}. \label{eq:eq2F1c}
\end{gather}
\end{subequations}
 These expressions, where the subscript '0' refers to the initial state of the respective variables, establish an independent equilibrium for each fluid: in the case of the neutral fluid, the force of gravity is balanced by the pressure gradients alone, while in the case of the charged fluid the magnetic pressure gradients also contribute to this balance.

In the 1F approach, the equilibrium cannot be established independently for each component of the plasma and all the forces affecting the charged and the neutral species have to be taken into account in a single equilibrium expression.
The equivalent condition in 1F model requires $\vec{u} = \vec{w}_{\rm{1F,0}} = \vec{0}$. By setting this condition in Eq. (\ref{eq:drift_whole}) one obtains
\begin{equation}
 [- \xi_{\rm{n,0}} \nabla \cdot \hat{p}_{\rm{m,0}} - \vec{G}_{\rm{0}}] =0.
\end{equation}
The same can be also verified, if Eqs. \eqref{eq:eq2F1n} and \eqref{eq:eq2F1c} are directly subtracted. Given the definition of $\vec{w}_{\rm{1F}}$ in Eq. \eqref{eq:drift_eq}, the above relation is equivalent to defining an ``equilibrium'' drift velocity, which is zero by definition,
\begin{equation}
    \vec{w}_{\rm{1F,0}} = \tilde{\eta}_{\rm{A,0}} [- \xi_{\rm{n,0}} \nabla \cdot \hat{p}_{\rm{m,0}} - \vec{G}_{\rm{0}}] = 0.
    \label{eq:w1F0}
\end{equation} 
Then it follows,
\begin{equation}
    \vec{w}_{\rm{amb,0}} = - \vec{w}_{\rm{G,0}}^{\rm{ideal}},
    \label{eq:Gamb0}
\end{equation}
which shows that in this kind of equilibrium the magnetic force is balanced by the thermal pressure function. The latter statement is in agreement with \cite{Bra1965aa}, where it is mentioned that the ambipolar and $\vec{G}$ terms are expected to balance each other for slow phenomena. Equation (\ref{eq:Gamb0}) shows that the inclusion of the $\vec{G}$ term in the 1F model is fundamental to describe an initial state consistent with the 2F equilibrium conditions given by Eqs. (\ref{eq:eq2F1n}) and (\ref{eq:eq2F1c}). If the thermal pressure function is not taken into account, Eq. (\ref{eq:drift_eq}) produces a drift velocity $\vec{w}_{\rm{1F,0}} \neq \vec{0}$ that would lead to the evolution of the equilibrium. 

The inspection of the second and third columns of Fig. \ref{fig:Driftmapz} reveals some areas (specially those close to the top and bottom boundaries of the domain) that have not been yet strongly affected by the turbulent phase of the RTI and where the terms $w_{\rm{amb,z}}$ and $w_{\rm{G,z}}^{\rm ideal}$ clearly have opposite signs but the same order of magnitude. This behaviour might be related to the evolution of the initial equilibrium. The first available snapshot from the RTI simulation series already presents weak non-zero velocities in regions not affected by the RTI (not shown here). 

\begin{figure}[t]
\centering
\includegraphics[width=0.7\hsize]{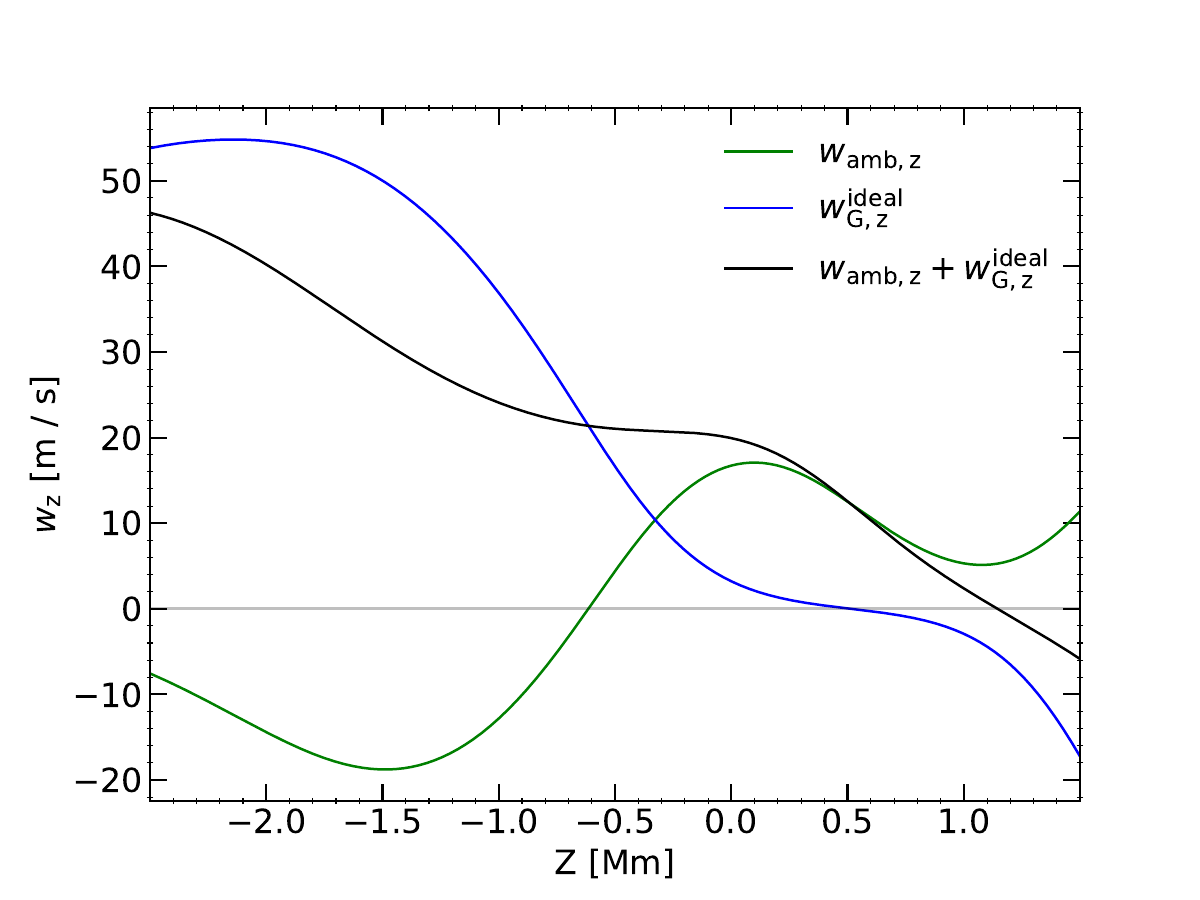}
\caption{Vertical component of the drift velocity due to the ambipolar term (green lines) and the thermal pressure function (blue lines) for a vertical column from the RTI simulations. Only the lines for P simulation are represented as the initial state of L2 is practically the same. 
The black line shows the total drift velocity due to the sum of both terms. }
\label{fig:Initital}
\end{figure}

Therefore, it is interesting to verify to which extent Eq. (\ref{eq:Gamb0}) is fulfilled for the initial equilibrium state of the RTI model. Note that the equilibrium used in those simulations, defined in  \cite{LeaDeVTha2014aa, PopLukKho2021aa}, was derived as a 1F equilibrium. We have computed $\vec{w}_{\rm{amb,0}}$ and  $\vec{w}_{\rm{G,0}}^{\rm{ideal}}$, in a vertical cut $x = 0$ through the simulation snapshot at the initial available time (which is about 3 second after the start of the evolution), see Fig. \ref{fig:Initital}. Although for most of the range of heights $w_{\rm{amb},z}$ and $w_{\rm{G,z}}^{\rm{ideal}}$ have opposite signs (in agreement with the discussion above), in the region $z \in \left[-0.5,0.5\right] \ \rm{Mm}$ they are both positive. In addition, the summation of the two contributions does not result in a zero net drift velocity. Therefore, this background model does not strictly fulfill Eq. (\ref{eq:Gamb0}). Nonetheless, the magnitude of the drift velocity is $|w_{\rm{z}}| \leq 50 \ \rm{m \ s^{-1}}$, much smaller than the values found in the nonlinear stage of the RTI (of the order of $1 \ \rm{km \ s^{-1}}$, as shown in Fig. \ref{fig:Driftmapz}) or the flow velocities of each fluid (which are of the order of $10^1 \ \rm{km \ s^{-1}}$ according to \cite{PopLukKho2021ab}). Hence, it is reasonable to say that Eq. (\ref{eq:Gamb0}) approximately holds for this initial configuration, that is, the thermal pressure function approximately balances the magnetic force. Such a small deviation from an equilibrium would not have a strong impact on the evolution of the instability and on the results and conclusions previously presented.

%%%%%%%%%%%%%%%%%%%%%%%%%%%%%%%%%%%%%%%
\section{Conclusions} \label{sec:6}
Throughout this work, the contributions of both the elastic and inelastic collisions into kinetic decoupling of charges and neutrals in partially ionized solar plasmas were discussed under SE and NEQ conditions. We started from a 2F description of a pure hydrogen plasma composed by a neutral and a charged fluid that are able to exchange momentum through hard sphere, CX and ionization/recombination collisions. In weakly decoupling conditions, the drift velocity between the fluids can be modeled in terms of the forces that act over each fluid. Among the terms contributing to the drift velocity is the ambipolar term. If inelastic collisions are taken into account, we obtain two corrections to the standard expression of ambipolar diffusion (see e.g. \cite{KhoColDia2014aa} or \cite{BalAleCol2018aa}): ionization/recombination rates modify the ambipolar diffusion coefficient and an additional contribution to the drift velocity appears if SE is not fulfilled, see Eq. (\ref{eq:drift_eq}). In order to estimate the magnitude of the corrections, we analyzed two physical scenarios: 1) the VAL3C quiet Sun model with hydrogen populations computed in SE by MULTI for a (5 + 1 levels) hydrogen model, and 2) 2F simulations of the RTI in a solar prominence thread, where populations are out of SE. From these experiments, we conclude that:
\begin{itemize}
    \item For typical conditions of the quiet Sun and under the assumption of SE, inelastic collisions play a secondary role in comparison with the elastic ones in the expression for the ambipolar coefficient, so the usual expression can be safely used.

    \item The RTI simulations show that the charge-neutral coupling of the plasma inside the prominence thread is dominated by elastic collisions, but in the hotter coronal region surrounding the thread the correction to the ambipolar coefficient gains importance and reaches some 10\%.
    
    \item The contribution of the NEQ drift is negligible in the RTI simulations studied here. 

    \item The drift velocity in the RTI simulations is dominated by the ideal part of the thermal pressure function, both in P and L2 cases. To the best of our knowledge, it is the first time that it is noted. Also, L2 simulation shows a small region dominated by the viscous drift. In the L2 simulation, where the magnetic forces gain importance, the ambipolar contribution to the drift velocity becomes comparable to the thermal pressure function one, and it is larger than in the P case where the evolution is purely hydrodynamical. 
    
    \item In the 2F equilibrium, when each of the fluids maintain its own force balance (see Eqs. \eqref{eq:eq2F1n} and \eqref{eq:eq2F1c}), the ambipolar and the thermal pressure function exactly balance each other, leading to drifts with the same magnitude but the opposite sign. In such a situation, thermal pressure function cannot be neglected compared to the ambipolar term. Conditions close to the self-balance are observed at the initial state and in the coronal environment of the snapshot analysed in this work, where deviations are mainly due to the ionisation/recombination imbalance.
    
\end{itemize}

For the sake of simplicity, many multi-fluid studies, such as \cite{Bra1965aa}, only consider elastic collisions. For example, \cite{VraKrs2013aa}, \cite{KhoColDia2014aa} and \cite{BalAleCol2018aa} have done very accurate discussions about the effects of elastic collisions in kinetic coupling and viscosity coefficients, but the inelastic ones were left out of the picture. In the work \cite{Mei2011aa} the authors proposed a 2F model that includes ionization/recombination rates that are independent of the radiation field, leading to the earlier efforts of \cite{LeaLukLin2012aa} to introduce them in a solar context. Although this description is a good approximation for coronal studies, where plasma is optically thin, it should be applied with caution to study the chromosphere, where radiation plays the major role.
Even so, the description by \cite{Mei2011aa, LeaLukLin2012aa} allows for a good approximation to understanding the role that ionization and recombination play in plasma dynamics, allowing to directly explore NEQ effects. 

Regarding the above, in this paper we found new insights into the mathematical description for the collisional coupling. If inelastic collisions are taken into consideration, the ambipolar diffusion coefficient is smaller than the classical one, which is in agreement with the conclusions of \cite{Bal2019aa}. Similarly, \cite{MurHilSno2022aa} found the coupling to increase in magnetic reconnection events. However, \cite{ZhaPoeLan2021aa} concluded that inelastic collisions favours decoupling. A possible explanation is that ionization/recombination processes enhance the cooling after the shock pass or reduce the density, reducing the coupling as consequence. 
We can be certain that the correction we propose to the ambipolar coefficient and the drift velocity acts towards increasing the coupling. However, since ionization/recombination influences thermodynamic conditions of the atmosphere, as well as the energy balance (not considered in this work), its overall effect is not straightforward to predict in realistic conditions, when all the factors are taken into account. We can speculate that shock waves and reconnection process seem to be the dynamical events where the inelastic corrections found in this work can certainly play a role.

The purpose of our work was not to encourage the use of either 1F or 2F models for a particular application. Rather, it is interesting to note that the comparison between the models allows to disentangle the role of different processes in the decoupling of the plasma components. In the 2F model, collisional terms are explicit, but the nature of the forces decoupling plasma components is not obvious. Unlike that, the equation for drift velocity, needed in the 1F modelling, shows more straightforwardly if magnetic, thermal, or other forces decouple the plasma. For example, such kind of analysis helped reveal the role of the thermal pressure function in the decoupling in the RTI simulations. While 1F model is much easier to apply for a global and realistic modeling and it is implemented in several well-known codes, a 2F model better describes some of the regimes studied in our paper, as, e.g., the environment surrounding cool structures in the solar corona (prominences, coronal rain). 

To systematically and consistently take into account radiation effects on MF plasma is extremely challenging not only technically, but also conceptually as the radiation effectively couples the individual fluids in a non-local, non-linear, and non-equilibrium manner. The state-of-the-art 1F codes include the effects of radiation in the rate equations \cite{LeeCarHan2007aa}. Recently, \cite{SnoDruHil2023aa} included the background radiation in the computation of the radiative rates in the MF approach. In future works, both kinds of collisions must be discussed together and a more realistic radiative transport must be included to deepen our understanding of the interaction between matter and radiation.

\ack{This work was supported by the Spanish Ministry of Science through the project PID2021-127487NB-I00 and by the International Space Science Institute (ISSI) in Bern, through ISSI International Team project 457: The Role of Partial Ionization in the Formation, Dynamics and Stability of Solar Prominences. It contributes to the deliverable identified in FP7 European Research Council grant agreement ERC-2017-CoG771310-PI2FA for the project “Partial Ionization: Two-fluid Approach”. M.M.G.M. and D.M. acknowledge support from the Spanish Ministry of Science and Innovation through the grant CEX2019-0000920-S of the Severo Ochoa Program. The author(s) wish to acknowledge the contribution of Teide High-Performance Computing facilities to the results of this research. TeideHPC facilities are provided by the Instituto Tecnológico y de Energías Renovables (ITER, SA). URL: http://teidehpc.iter.es. We also acknowledge Beatrice Popescu Braileanu for sharing her simulations with us and all the helpful comments about them, as well as Peter Hunana and Jacob Heerikhuisen for their useful comments about collision physics.}

\section{Figures \& Tables}

%%%%%%%%%%%%%%%%%%%%%%%%%%%%%%%%%%%%%%%%%%%%%%%%%%%%%%%%%%%%%%%%%%%
%%%%%%%%%%%%%%%%%%%%%%%%%%%%%%%%%%%%%%%%%%%%%%%%%%%%%%%%%%%%%%%%%%%

%%%%%%%%%%%%%%%%%%%%%%%%%%%%%%%%%%%%%%%%%%%%%%%%%%%%%%%%%%%%%%%%%%%
%%%%%%%%%%%%%%%%%%%%%%%%%%%%%%%%%%%%%%%%%%%%%%%%%%%%%%%%%%%%%%%%%%%

%%%%%%%%%%%%%%%%%%%%%%%%%%%%%%%%%%%%%%%%%%%%%%%%%%%%%%%%%%%%%%%%%%%
%%%%%%%%%%%%%%%%%%%%%%%%%%%%%%%%%%%%%%%%%%%%%%%%%%%%%%%%%%%%%%%%%%%

%%%%%%%%%%%%%%%%%%%%%%%%%%%%%%%%%%%%%%%%%%%%%%%
%%%%%%%%%%%%%%%%%%%%%%%%%%%%%%%%%%%%%%%%%%%%%%%%%

%%%%%%%%%%%%%%%%%%%%%%%%%%%%%%%%%%%%%%%%%%%%%%%%%
%%%%%%%%%%%%%%%%%%%%%%%%%%%%%%%%%%%%%%%%%%%%%%%%%

%%%%%%%%%%%%%%%%%%%%%%%%%%%%%%%%%%%%%%%%%%%%%%%
%%%%%%%%%%%%%%%%%%%%%%%%%%%%%%%%%%%%%%%%%%%%%%%%%

%%%%%%%%%%%%%%%%%%%%%%%%%%%%%%%%%%%%%%%%%%%%%%%
%%%%%%%%%%%%%%%%%%%%%%%%%%%%%%%%%%%%%%%%%%%%%%%%%

%%%%%%%%%%%%%%%%%%%%%%%%%%%%%%%%%%%%%%%%%%%%%%%
%%%%%%%%%%%%%%%%%%%%%%%%%%%%%%%%%%%%%%%%%%%%%%%%%

\begin{appendix}

\section{Comments on cross-sections\label{app:cross}}
In this appendix, we briefly discuss the concept of cross section. The differential cross-section $d\sigma/d\Omega$ accounts for the fraction of particles coming from the differential area $d\sigma$ of the collisional plane scattered into a solid angle $d\Omega$ \cite{SchNag2009aa}. In a practical sense, considering a flux of particles $\Gamma$ in their way to collide with the target particle, the number of particles scattered into a solid angle $d\Omega$ is
\begin{equation}
    dN_{\rm{\alpha}} = \frac{d\sigma}{d\Omega}(g_{\rm{\alpha\beta}}, \theta) d\Omega \Gamma,
    \label{eq:diffcs}
\end{equation}
being $g_{\rm{\alpha\beta}} = |\vec{v}_{\rm{\alpha}} - \vec{v}_{\rm{\beta}}|$ the relative microscopic velocity of the colliding particles and $\theta$ the scattering angle.

A useful quantities are the moments of the differential cross-section. The most used ones in fluid dynamics are defined as follows \cite{KrsSch1999aa}:
\begin{subequations}
    \begin{gather}
        \sigma_{\rm{0}} = 2\pi\int_0^\pi d\theta \sin \theta \frac{d \sigma}{d \Omega}, \label{eq:sigma0}\\
        \sigma_{\rm{1}} = 2\pi\int_0^\pi d\theta \sin \theta (1 - \cos \theta) \frac{d \sigma}{d \Omega}, \label{eq:sigma1}\\
        \sigma_2 = 2\pi\int_0^\pi d\theta \sin^3 \theta \frac{d \sigma}{d \Omega}. \label{eq:sigma2}
    \end{gather}
\end{subequations}

They are called scattering, momentum transfer and viscosity cross-sections, respectively, and account for the amount of particles scattered or momentum and energy exchanged during the collision. The one of our concern is $\sigma_1$, which is usually referred in literature simply as $\sigma$. In the fluid approach, it is usually necessary to evaluate the average value of this cross-section, e.g., the quantity $\langle\sigma g_{\rm{\alpha\beta}} \rangle$ for the motion equation. For example, for hard sphere collisions, $\sigma \propto g_{\rm{\alpha\beta}}^{-1}$ and an analytic expression can be found \cite{Bra1965aa, Dra1986aa}. As in most of the cases the integrals above are hard to solve, it is common to take the average of $\sigma(g_{\rm{\alpha\beta}})$ with a Maxwelian distribution as an approximation to the cross-section, so $\sigma(g_{\rm{\alpha\beta}})\approx \tilde{\sigma}(T)$ (e.g. see \cite{SchNag2009aa}).

\end{appendix}

%\aareferences

\end{document}